\documentclass{pasj00}

\Received{$\langle$reception date$\rangle$}
\Accepted{$\langle$acception date$\rangle$}
\Published{$\langle$publication date$\rangle$}

\SetRunningHead{Astronomical Society of Japan}{Usage of \texttt{pasj00.cls}}

\usepackage{times}

\begin{document}

\title{Rotation Modulations and Distributions of The Flare Occurrence Rates \\
On The Surface Of Five UV Ceti Type Stars}

\author{H. A. Dal and S. Evren}
\affil{Department Of Astronomy and Space Sciences, University of Ege, \\
Bornova, 35100 ~\.{I}zmir, Turkey}

\email{ali.dal@ege.edu.tr}


\maketitle

\begin{abstract}

In this study, we discuss stellar spots, stellar flares and also the relation between these two magnetic proccess that take place on UV Ceti stars. In addition, the hypothesis about slow flares described by \citet{Gur86} will be discussed. All these discussions are based on the results of three years of observations of the UV Ceti type stars AD Leo, EV Lac, V1005 Ori, EQ Peg and V1054 Oph. First of all, the results show that the stellar spot activity occurs on the stellar surface of EV Lac, V1005 Ori and EQ Peg, while AD Leo does not show any short-term variability and V1054 Oph does not exhibits any variability. We report new ephemerides, for EV Lac, V1005 Ori and EQ Peg, obtained from the time series analyses. The phases, computed in intervals of 0.10 phase length, where the mean flare occurence rates get maximum amplitude, and the phases of rotational modulation were compared to investigate whether there is any longitudinal relation between stellar flares and spots. Although, the results show that flare events are related with spotted areas on the stellar surfaces in some of the observing seasons, we did not find any clear correlation among them. Finally, it is tested whether slow flares are the fast flares occurring on the opposite side of the stars according to the direction of the observers as mentioned in the hypothesis developed by \citet{Gur86}. The flare occurence rates reveal that both slow and fast flares can occur in any rotational phases. The flare occurence rates of both fast and slow flares are varying in the same way along the longitudes for all program stars. These results are not expected based on the case mentioned in the hypothesis.

\end{abstract}

\section{Introduction}

Many samples of UV Ceti type stars posses the stellar spot activity known as BY Dra Syndrome. The BY Dra Syndrome among the UV Ceti stars was first found by \citet{Kro52}. He reported the existence of sinusoidal-like variations at out-of-eclipses of the eclipsing binary star YY Gem. \citet{Kro52} explained this sinusoidal-like variation at out-of-eclipse as a heterogeneous temperature on star surface, which was called BY Dra Syndrome by \citet{Kun75}. This interpretation of BY Dra Syndrome in terms of dark regions of the surface of rotating stars was confirmed, based on more rigorous arguments, by later works of \citet{Fer73, Bop73, Vog75, Fri75}.

Since the most of solar flares occur over solar spot regions, in the stellar case it is also expected to find a correlation between the frequency of flares and the effects caused by spots in the light curve. In order to determine a similar relation among the stars, lots of studies have been made using UV Ceti type stars showing stellar spot activity such as BY Dra. One of these studies was reported by \citet{Bop74} on YY Gem. \citet{Bop74} did not find a clear correlation between the location and extent in longitude of flares and spots on YY Gem. Moreover, he notes that the longitudinal extent he derived for a flare-producing region is in good agreement with the longitudinal extents of starspots previously calculated for BY Dra and CC Eri by \citet{Bop73}. In another study, \citet{Pet83} compared longitudes of the stellar spots obtained from two years observations of YZ CMi and EV Lac with longitudes of flare events and distributions of flare energy and frequencies along the longitude obtained in the same work. Direct comparisons and statistical tests are not able to reveal positive relationships between flare frequency or flare energy and the position of the spotted region. In another extended work, \citet{Let97} looked for whether there is any relation between stellar spots and flares observed from 1967 to 1977 in the observations of EV Lac. The authors were able to find a relation in the year 1970. They could not find any relation in other observing seasons because of the higher threshold of the system used for flare detection. In the last years, \citet{Gar03} found some flares occurring in the same active area with other activity patterns with using simultaneous observations.

Since no correlation is found between stellar spots and flares, a hypothesis about fast and slow flares was put forward. The hypothesis is based on the work named as Fast Electron Hypothesis. According to this hypothesis, the shape of a flare light variation depends on the location of the event on the star surface in respect to direction of observer. If the flaring area is on the front side of the star according to the observer, the light variation shape looks as a fast flare. If the flaring area is on the opposite side of the star according to the observer, the light variation shape looks as a slow flare \citep{Gur65, Gur86}. In addition, \citet{Gur88} described two types of flares to model flare light curves. \citet{Gur88} indicated that thermal processes are dominant in the processes of slow flares, which are 95$\%$ of all flares observed in UV Ceti type stars. Non-thermal processes are dominant in the processes of fast flares, which are classified as "other" flares. According to \citet{Gur88}, there is a large energy difference between these two types of flares. Moreover, \citet{Dal10} developed a rule to the classifying of fast and slow flares. When the ratios of flare decay times to flare rise times are computed for two types of flares, the ratios never exceed 3.5 for all slow flares. On the other hand, the ratios are always above 3.5 for fast flares. It means that if the decay time of a flare is 3.5 times longer than its rise time at least, the flare is a fast flare. If not, the flare is a slow flare.

In this paper, the results obtained from Johnson UBVR observations of AD Leo, EV Lac and V1005 Ori will be discussed. \citet{Sha74} and \citet{Bop77} reported that V1005 Ori is a flare star that exhibits rotational modulation due to stellar spots. The authors found an amplitude variation of $0^{m}.08$ with a period of $1^{d}.96$ in V band. Besides, \citet{Bop78} examined photometric data in the time series analyses and found 4 periods for rotational modulation. The period of $1^{d}.858$ is suggested as the most probable period among them. On contrary, in B band observations of 1981, a $4.56\pm0.01$ day period variation with an amplitude of $0^{m}.16$ was found \citep{Byr84}. No important light curve changes are seen in the years 1996 and 1997, while the minimum phases of rotation modulation is varied from $0^{P}.40$ to $0^{P}.55$. The amplitude of the curves is $0^{m}.10$ in the year 1996, but in the year 1997 it gets larger than the previous ones \citep{Ama01}.

In the case of AD Leo, it is a debate issue whether AD Leo has any stellar spot activity, or not. \citet{Chu74} and \citet{Mul74} show that AD Leo does not exhibit any rotational modulation caused by stellar spots. Besides, \citet{And79} found no variations at the $0.02$ magnitude level during the period 1978, May 10 to 17. However, \citet{Spi86} reveals that AD Leo demonstrates BY Dra Syndrome with a period of $2.7\pm0.05$ days. In addition, \citet{Pan93} confirmed this period of AD Leo for BY Dra variation.

On the other hand, EV Lac is a well known active star with both high level flare and stellar spot activities. \citet{Mah81} indicated that the star has no variation caused by rotational modulation in the observations in B band from 1972 to 1976. However, \citet{Pet80} showed a rotational modulation with a period of $4^{d}.378$ and an amplitude of $0^{m}.07$. \citet{Pet83}, based on continuous observations from 1979 to 1981, renewed the ephemerides of the variation as a period of $4^{d}.375$ and an amplitude of $0^{m}.08$. This indicates that the light curves of EV Lac were almost constant for 2.5 years because the spot groups on the star are stable during these 2.5 years. Using the renewed ephemerides, \citet{Kle87} found the amplitude of light curve enlarging from $0^{m}.08$ to $0^{m}.16$ in the year 1986. On the other hand, no variation was seen in the light curve of the year 1987. \citet{Pet92} showed that the spotted area is located in the same semi-sphere on EV Lac for 10 years. Comparing the phases of the light curve minima caused by rotational modulation with the flare frequencies and the distribution of the flare equivalent durations for YZ CMi and EV Lac, \citet{Pet83} showed that there is no relation between the flare activity and stellar spot activity on these stars.

EQ Peg is classified as a metal-rich star and it is a member of the young disk population in the galaxy \citep{Vee74, Fle95}. EQ Peg is a visual binary \citep{Wil54}. Both components are flare stars \citep{Pet83}. Angular distance between components is given as a value between 3$\arcsec$.5 and 5$\arcsec$.2 \citep{Hai87, Rob04}. One of the components is 10.4 mag and the other is 12.6 mag in V band \citep{Kuk69}. Observations show that flares on EQ Peg generally come from the fainter component \citep{Fos95}. \citet{Rod78} proved that 65$\%$ of the flares come from faint component and about 35$\%$ from the brighter component.

The fourth star in this study is V1054 Oph, whose flare activity was discovered by \citet{Egg65}. \citet{Nor07} demonstrated that EQ Peg has a variability with the period of $1^{d}.0664$. V1054 Oph (= Wolf 630ABab, Gliese 644ABab) is a member of Wolf star group \citep{Joy47, Joy74}. Wolf 630ABab, Wolf 629AB (= Gliese 643AB) and VB8 (= Gliese 644C), are the members of the main triplet system, whose scheme is shown in Fig.1 given by \citet{Pet84}. The masses were derived for each components of Wolf 630ABab by \citet{Maz01}. The author showed that the masses are 0.41 $M_{\odot}$ for Wolf 629A, 0.336 $M_{\odot}$ for Wolf 630Ba and 0.304 $M_{\odot}$ for Wolf 630Bb. In addition, \citet{Maz01} demonstrated that the age of the system is about 5 Gyr.

In this study, for each program stars, we analyse the variations at out-of-flare for each light curves obtained in Johnson UBVR observations, or not. Although all of them show high flare activity, EV Lac, V1005 Ori and EQ Peg exhibit stellar spot activity. On the other hand, the spot activity is not obvious for AD Leo. It is discussed whether AD Leo has any stellar spot activity, or not. Finally, this work do not demonstrate any variation from rotational modulations. To perform this kind of studies we would require a long term observing program. As a part of this study, the phase distributions for both fast and slow flares are examined in terms of the minimum phases of rotational modulation. Thus, hypothesis developed by \citet{Gur86} is tested.

\section{Observations and Analyses}

\subsection{Observations}

The observations were acquired with a High-Speed Three Channel Photometer attached to the 48 cm Cassegrain type telescope at Ege University Observatory. Observations were grouped in two schedules. Using a tracking star in second channel of the photometer, flare observations were only continued in standard Johnson U band with exposure times between 2 and 10 seconds. The same comparison stars were used for all observations. The second observation schedule was used for determining whether there was any variation out-of-flare. Pausing flare patrol of program stars, we observed them once or twice a night, when they were close to the celestial meridian. Using a tracking star in second channel of the photometer, the observations in this schedule were made with the exposure time of 10 seconds in each band of standard Johnson UBVR system, respectively. There were any delay between the exposure in different filters due to the High-Speed Three Channel Photometer.

Although the program and comparison stars are so close on the sky, differential atmospheric extinction corrections were applied. The atmospheric extinction coefficients were obtained from the observations of the comparison stars on each night. Moreover, the comparison stars were observed with the standard stars in their vicinity and the reduced differential magnitudes, in the sense variable minus comparison, were transformed to the standard system using procedures outlined by \citet{Har62}. The standard stars are listed in the catalogues of \citet{Lan83} and \citet{Lan92}. And also, the de-reddened colour of the systems were computed. Heliocentric corrections were also applied to the times of observations. The mean averages of the standard deviations are $0^{m}.015$, $0^{m}.009$, $0^{m}.007$ and $0^{m}.007$ for the observations acquired in standard Johnson UBVR bands, respectively. To compute the standard deviations of observations, we use the standard deviations of the reduced differential magnitudes in the sense comparisons (C1) minus check (C2) stars for each night. There is no variation in the standard brightness comparison stars.

\begin{table*}
\begin{center}
\caption{Basic parameters for the stars studied and their comparison (C1) and check (C2) stars.\label{tbl-1}}
\begin{tabular}{lccc}
\hline\hline
\textbf{Stars} & \textbf{V (mag)} & \textbf{B-V (mag)} & \textbf{Spectral Type}  \\
\hline 
\textbf{AD Leo} & 9.388 & 1.498 & M3 \\
C1 = HD 89772 & 8.967 & 1.246 & K6-K7 \\
C2 = HD 89471 & 7.778 & 1.342 & K8 \\
\hline 
\textbf{V1005 Ori} & 10.090 & 1.307 & K7 \\
C1 = BD+01 870 & 8.800 & 1.162 & K5 \\
C2 = HD 31452 & 9.990 & 0.920 & K2 \\
\hline 
\textbf{EV Lac} & 10.313 & 1.554 & M3 \\
C1 = HD 215576 & 9.227 & 1.197 & K6 \\
C2 = HD 215488 & 10.037 & 0.881 & K1-K2 \\
\hline
\textbf{EQ Peg} & 10.170 & 1.574 & M3-M4 \\
C1 = SAO 108666 & 9.598 & 0.745 & G8 \\
C2 = SAO 91312 & 9.050 & 1.040 & K3-K4 \\
\hline
\textbf{V1054 Oph} & 8.996 & 1.552 & M3 \\
C1 = HD 152678 & 7.976 & 1.549 & M3 \\
C2 = SAO 141448 & 9.978 & 0.805 & K0 \\
\hline
\end{tabular}
\end{center}
\end{table*}

The identities of all programme stars and their comparisons are given in Table 1. All the magnitudes and colour indexes in the table were were taken from \citet{Dal10} and \citet{Dal11}.

\begin{table*}
\begin{center}
\caption{Observational reports of the program stars for each observing season.\label{tbl-2}}
\begin{tabular}{lcccc}
\hline\hline
\textbf{Stars} & \textbf{Observing} & \textbf{HJD Interval} & \textbf{Filter} & \textbf{Number} \\
& \textbf{Season} & \textbf{(+24 00000)} & & \textbf{of Night} \\
\hline 
\textbf{AD Leo} & 2004/2005 & 53377-53514 & $UBVR$ & 15 \\
& 2005/2006 & 53717-53831 & $UBVR$ & 16 \\
& 2006/2007 & 54071-54248 & $UBVR$ & 13 \\
\hline 
\textbf{EV Lac} & 2004 & 53202-53240 & $UBVR$ & 11 \\
& 2004 & 53260-53312 & $UBVR$ & 15 \\
& 2005 & 53554-53652 & $UBVR$ & 18 \\
& 2006 & 53863-54058 & $UBVR$ & 27 \\
\hline 
\textbf{V1005 Ori} & 2004/2005 & 53353-53453 & $UBVR$ & 11 \\
& 2005/2006 & 53640-53812 & $UBVR$ & 22 \\
& 2006/2007 & 53984-54158 & $UBVR$ & 15 \\
\hline
\textbf{EQ Peg} & 2004 & 53236-53335 & $UBVR$ & 13 \\
& 2005 & 53621 - 53686 & $UBVR$ & 9 \\
\hline 
\textbf{V1054 Oph} & 2004 & 53136-53167 & $UBVR$ & 9 \\
& 2005 & 53505 - 53564 & $UBVR$ & 9 \\
& 2006 & 53861 - 53894 & $UBVR$ & 7 \\
\hline
\end{tabular}
\end{center}
\end{table*}

The observation reports of programme stars are given in Table 2. In this table, "observing seasons" are given with HJD intervals. The observing season refer the period of the year in which each star can be seen from the site of observation. In the last column of the table, the number of night refers to total number of nights dedicated to observe the corresponding star in a given observing season.

\subsection{The Time Series Analyses}

All data sets were analysed with the method of Discrete Fourier Transform (DFT) \citep{Sca82}. The results obtained from DFT were tested by two other methods. One of them is CLEANest, which is another Fourier method \citep{Fos95}, and the second method is the Phase Dispersion Minimization (PDM), which is a statistical method \citep{Ste78}. These methods confirmed the results obtained by DFT. Although analyses showed a variation in each observation season for EV Lac, V1005 Ori and EQ Peg, analyses do not indicate any variation for both AD Leo and V1054 Oph in each season. All data set obtained in Johnson UBVR bands are given in Table 3 for each star. The star names are given in the first column, the observing seasons are given in the second column and HJDs are in the third column. V band magnitudes, U-B, B-V and V-R colour indexes are given in the next four columns. The number between brackets are the standard deviations in the columns.

The values of U-B colour indexes listed in Table 3 were not use in time series analyses, because the U-B colour indexes are much more sensitive to the flare activity on the surface of the programme stars. When a small flare occurred to detected in respect to the higher threshold, it is not seen any sign in V light, B-V and V-R colours. On the other hand, some distinctive sign is seen in U band light and U-B colour.

\begin{figure*}
\hspace{15mm}
\vspace{5mm}
\FigureFile(155mm,60mm){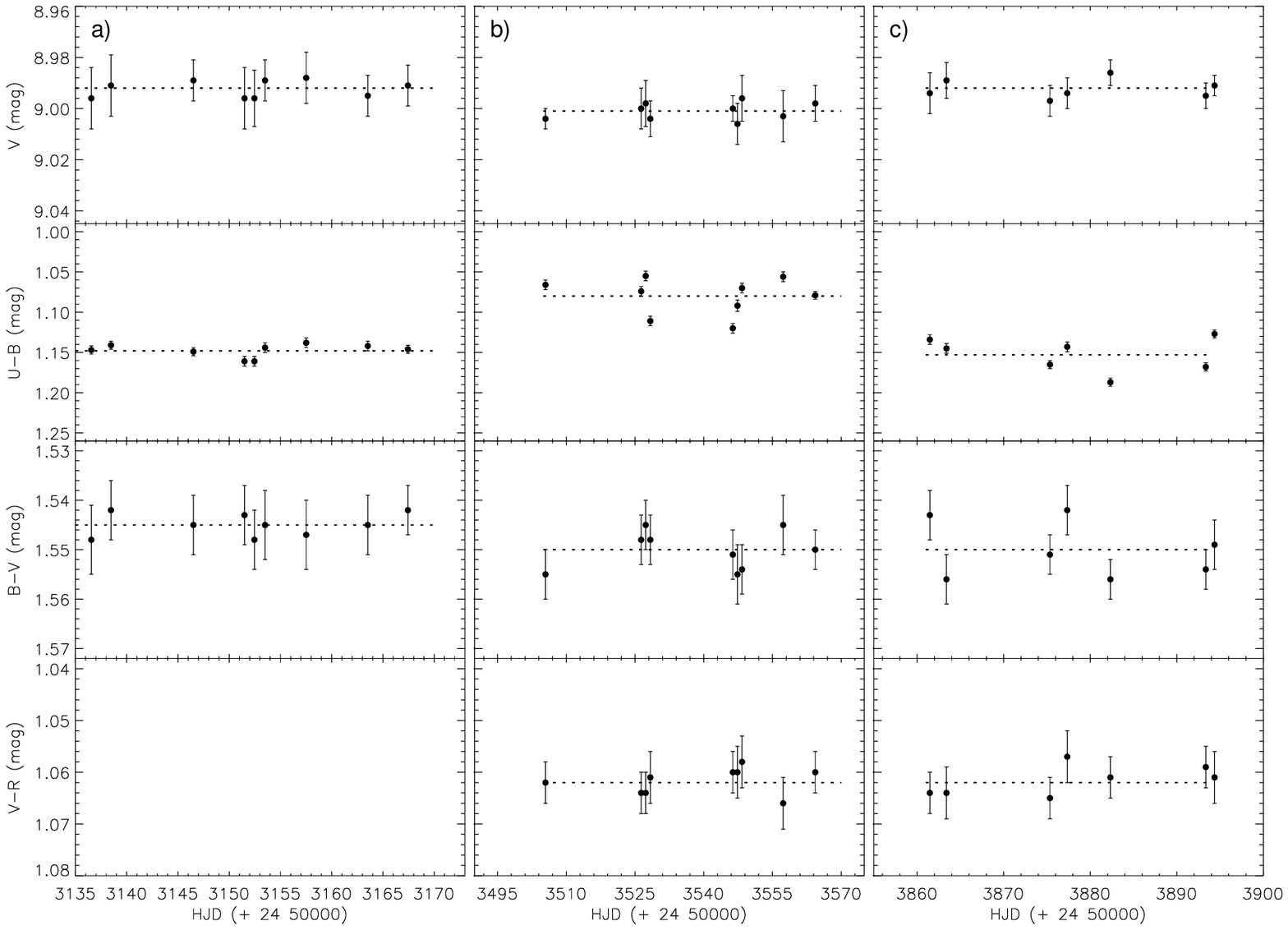}
\caption{The light and colour curves obtained from the observations of V1054 Oph in this study are seen for the seasons 2004 (a), 2005 (b) and 2006 (c). Dashed lines represent the mean brightness and colour level obtained from the time series analyses. \label{fig1}}
\end{figure*}

\begin{longtable}{lcccccc}
\caption{All observation data obtained in Johnson UBVR bands for all program stars. The number between brackets are the standard deviations in the columns.\label{tbl-3}}
\hline
Star	&	Observing	&	HJD	&	V	&	U-B	&	B-V	&	V-R	\\
	&	Season	&	(+24 00000)	&	(mag)	&	(mag)	&	(mag)	&	(mag)	\\
\hline\hline
AD Leo 	&	 2004/2005 	&	 53377.4886(0.0055) 	&	 9.402(0.001) 	&	 0.957(0.009) 	&	 1.502(0.005) 	&	 1.124(0.004) \\
AD Leo 	&	 2004/2005 	&	 53381.4822(0.0047) 	&	 9.411(0.008) 	&	 0.936(0.015) 	&	 1.486(0.007) 	&	 1.127(0.005) \\
AD Leo 	&	 2004/2005 	&	 53412.4029(0.0008) 	&	 9.380(0.003) 	&	 1.004(0.019) 	&	 1.492(0.001) 	&	 1.134(0.001) \\
AD Leo 	&	 2004/2005 	&	 53413.4187(0.0045) 	&	 9.412(0.002) 	&	 0.869(0.015) 	&	 1.484(0.003) 	&	 1.136(0.001) \\
AD Leo 	&	 2004/2005 	&	 53444.3612(0.0044) 	&	 9.424(0.002) 	&	 0.997(0.006) 	&	 1.500(0.005) 	&	 1.129(0.004) \\
AD Leo 	&	 2004/2005 	&	 53446.2663(0.0048) 	&	 9.399(0.001) 	&	 0.983(0.011) 	&	 1.497(0.003) 	&	 1.119(0.002) \\
AD Leo 	&	 2004/2005 	&	 53465.3122(0.0073) 	&	 9.404(0.009) 	&	 0.960(0.039) 	&	 1.497(0.011) 	&	 1.127(0.008) \\
AD Leo 	&	 2004/2005 	&	 53469.2653(0.0046) 	&	 9.379(0.003) 	&	 0.954(0.018) 	&	 1.519(0.002) 	&	 1.105(0.004) \\
AD Leo 	&	 2004/2005 	&	 53470.2777(0.0040) 	&	 9.408(0.003) 	&	 0.938(0.024) 	&	 1.510(0.005) 	&	 1.113(0.002) \\
AD Leo 	&	 2004/2005 	&	 53471.2669(0.0045) 	&	 9.404(0.005) 	&	 0.947(0.007) 	&	 1.496(0.007) 	&	 1.119(0.003) \\
AD Leo 	&	 2004/2005 	&	 53499.2888(0.0039) 	&	 9.405(0.003) 	&	 0.918(0.004) 	&	 1.494(0.003) 	&	 1.148(0.006) \\
AD Leo 	&	 2004/2005 	&	 53500.2896(0.0047) 	&	 9.380(0.008) 	&	 0.968(0.044) 	&	 1.500(0.007) 	&	 1.119(0.009) \\
AD Leo 	&	 2004/2005 	&	 53502.2878(0.0055) 	&	 9.396(0.004) 	&	 0.957(0.008) 	&	 1.501(0.005) 	&	 1.109(0.006) \\
AD Leo 	&	 2004/2005 	&	 53505.2925(0.0046) 	&	 9.395(0.003) 	&	 0.928(0.016) 	&	 1.502(0.004) 	&	 1.120(0.003) \\
AD Leo 	&	 2004/2005 	&	 53514.2952(0.0037) 	&	 9.394(0.011) 	&	 0.961(0.035) 	&	 1.500(0.005) 	&	 1.127(0.009) \\
AD Leo 	&	 2005/2006 	&	 53717.5037(0.0058) 	&	 9.404(0.001) 	&	 0.998(0.010) 	&	 1.506(0.001) 	&	 1.120(0.002) \\
AD Leo 	&	 2005/2006 	&	 53744.4914(0.0044) 	&	 9.399(0.003) 	&	 0.988(0.011) 	&	 1.498(0.004) 	&	 1.121(0.004) \\
AD Leo 	&	 2005/2006 	&	 53757.4158(0.0036) 	&	 9.387(0.001) 	&	 0.964(0.006) 	&	 1.501(0.001) 	&	 1.109(0.001) \\
AD Leo 	&	 2005/2006 	&	 53763.5181(0.0035) 	&	 9.385(0.002) 	&	 0.969(0.006) 	&	 1.500(0.003) 	&	 1.112(0.004) \\
AD Leo 	&	 2005/2006 	&	 53764.5537(0.0040) 	&	 9.400(0.002) 	&	 0.936(0.009) 	&	 1.494(0.003) 	&	 1.122(0.002) \\
AD Leo 	&	 2005/2006 	&	 53765.5831(0.0038) 	&	 9.389(0.002) 	&	 0.920(0.009) 	&	 1.504(0.002) 	&	 1.115(0.003) \\
AD Leo 	&	 2005/2006 	&	 53769.4748(0.0038) 	&	 9.396(0.002) 	&	 0.926(0.007) 	&	 1.499(0.001) 	&	 1.114(0.002) \\
AD Leo 	&	 2005/2006 	&	 53771.4091(0.0039) 	&	 9.380(0.003) 	&	 0.952(0.010) 	&	 1.506(0.003) 	&	 1.117(0.004) \\
AD Leo 	&	 2005/2006 	&	 53782.4414(0.0036) 	&	 9.395(0.002) 	&	 0.971(0.009) 	&	 1.529(0.002) 	&	 1.120(0.003) \\
AD Leo 	&	 2005/2006 	&	 53787.4247(0.0035) 	&	 9.394(0.003) 	&	 1.008(0.008) 	&	 1.497(0.003) 	&	 1.112(0.003) \\
AD Leo 	&	 2005/2006 	&	 53788.3733(0.0035) 	&	 9.391(0.002) 	&	 0.968(0.006) 	&	 1.486(0.003) 	&	 1.121(0.003) \\
AD Leo 	&	 2005/2006 	&	 53812.2796(0.0027) 	&	 9.385(0.003) 	&	 0.959(0.011) 	&	 1.500(0.003) 	&	 1.113(0.004) \\
AD Leo 	&	 2005/2006 	&	 53816.2875(0.0038) 	&	 9.386(0.002) 	&	 0.870(0.009) 	&	 1.502(0.004) 	&	 1.107(0.003) \\
AD Leo 	&	 2005/2006 	&	 53821.2994(0.0035) 	&	 9.381(0.008) 	&	 0.985(0.012) 	&	 1.494(0.007) 	&	 1.113(0.010) \\
AD Leo 	&	 2005/2006 	&	 53827.2550(0.0035) 	&	 9.385(0.003) 	&	 0.934(0.024) 	&	 1.505(0.005) 	&	 1.103(0.003) \\
AD Leo 	&	 2005/2006 	&	 53831.2530(0.0034) 	&	 9.383(0.001) 	&	 1.014(0.007) 	&	 1.520(0.004) 	&	 1.109(0.002) \\
AD Leo 	&	 2006/2007 	&	 54071.5791(0.0034) 	&	 9.379(0.005) 	&	 0.995(0.054) 	&	 1.495(0.006) 	&	 1.110(0.010) \\
AD Leo 	&	 2006/2007 	&	 54085.6298(0.0040) 	&	 9.378(0.007) 	&	 0.987(0.029) 	&	 1.488(0.006) 	&	 1.101(0.009) \\
AD Leo 	&	 2006/2007 	&	 54093.4845(0.0037) 	&	 9.376(0.005) 	&	 0.938(0.012) 	&	 1.501(0.012) 	&	 1.107(0.009) \\
AD Leo 	&	 2006/2007 	&	 54097.5495(0.0047) 	&	 9.376(0.009) 	&	 0.932(0.050) 	&	 1.491(0.007) 	&	 1.109(0.006) \\
AD Leo 	&	 2006/2007 	&	 54114.5302(0.0035) 	&	 9.375(0.009) 	&	 0.976(0.045) 	&	 1.500(0.012) 	&	 1.104(0.007) \\
AD Leo 	&	 2006/2007 	&	 54115.5539(0.0057) 	&	 9.369(0.005) 	&	 0.982(0.044) 	&	 1.498(0.008) 	&	 1.100(0.007) \\
AD Leo 	&	 2006/2007 	&	 54122.5021(0.0044) 	&	 9.372(0.008) 	&	 0.970(0.030) 	&	 1.500(0.008) 	&	 1.097(0.007) \\
AD Leo 	&	 2006/2007 	&	 54138.4519(0.0000) 	&	 9.381(0.006) 	&	 0.941(0.066) 	&	 1.496(0.005) 	&	 1.109(0.005) \\
AD Leo 	&	 2006/2007 	&	 54227.2863(0.0039) 	&	 9.374(0.005) 	&	 0.968(0.046) 	&	 1.497(0.010) 	&	 1.103(0.004) \\
AD Leo 	&	 2006/2007 	&	 54232.3069(0.0034) 	&	 9.372(0.003) 	&	 0.936(0.032) 	&	 1.492(0.010) 	&	 1.087(0.008) \\
AD Leo 	&	 2006/2007 	&	 54236.3290(0.0035) 	&	 9.368(0.006) 	&	 0.977(0.037) 	&	 1.482(0.007) 	&	 1.087(0.008) \\
AD Leo 	&	 2006/2007 	&	 54237.3158(0.0048) 	&	 9.363(0.009) 	&	 0.862(0.056) 	&	 1.486(0.011) 	&	 1.086(0.010) \\
AD Leo 	&	 2006/2007 	&	 54248.3084(0.0050) 	&	 9.356(0.004) 	&	 0.904(0.046) 	&	 1.484(0.007) 	&	 1.097(0.003) \\
\hline
EV Lac 	&	 2004-Set 1 	&	 53202.4799(0.0000) 	&	 10.271(0.007) 	&	 1.253(0.045) 	&	 1.577(0.010) 	&	 1.163(0.010) \\
EV Lac 	&	 2004-Set 1 	&	 53204.4809(0.0009) 	&	 10.336(0.006) 	&	 1.231(0.019) 	&	 1.564(0.008) 	&	 1.174(0.002) \\
EV Lac 	&	 2004-Set 1 	&	 53207.4523(0.0058) 	&	 10.251(0.008) 	&	 1.183(0.032) 	&	 1.573(0.011) 	&	 1.159(0.010) \\
EV Lac 	&	 2004-Set 1 	&	 53211.3915(0.0056) 	&	 10.276(0.007) 	&	 1.228(0.067) 	&	 1.553(0.017) 	&	 1.178(0.008) \\
EV Lac 	&	 2004-Set 1 	&	 53212.3637(0.0014) 	&	 10.265(0.010) 	&	 1.168(0.056) 	&	 1.590(0.017) 	&	 1.197(0.014) \\
EV Lac 	&	 2004-Set 1 	&	 53226.3835(0.0008) 	&	 10.320(0.010) 	&	 1.182(0.046) 	&	 1.561(0.012) 	&	 1.165(0.009) \\
EV Lac 	&	 2004-Set 1 	&	 53227.3451(0.0011) 	&	 10.304(0.007) 	&	 1.236(0.052) 	&	 1.582(0.012) 	&	 1.166(0.011) \\
EV Lac 	&	 2004-Set 1 	&	 53228.3398(0.0009) 	&	 10.278(0.013) 	&	 1.300(0.062) 	&	 1.576(0.019) 	&	 1.154(0.012) \\
EV Lac 	&	 2004-Set 1 	&	 53232.3968(0.0058) 	&	 10.293(0.014) 	&	 1.236(0.017) 	&	 1.558(0.010) 	&	 1.171(0.012) \\
EV Lac 	&	 2004-Set 1 	&	 53236.3282(0.0011) 	&	 10.292(0.003) 	&	 1.210(0.018) 	&	 1.574(0.005) 	&	 1.177(0.008) \\
EV Lac 	&	 2004-Set 1 	&	 53240.3283(0.0011) 	&	 10.325(0.011) 	&	 1.324(0.013) 	&	 1.564(0.017) 	&	 1.178(0.009) \\
EV Lac 	&	 2004-Set 2 	&	 53260.2571(0.0016) 	&	 10.298(0.007) 	&	 1.301(0.046) 	&	 1.574(0.006) 	&	 1.165(0.003) \\
EV Lac 	&	 2004-Set 2 	&	 53263.2966(0.0012) 	&	 10.300(0.010) 	&	 1.299(0.064) 	&	 1.558(0.010) 	&	 1.176(0.009) \\
EV Lac 	&	 2004-Set 2 	&	 53264.2561(0.0012) 	&	 10.281(0.017) 	&	 1.404(0.062) 	&	 1.571(0.019) 	&	 1.153(0.015) \\
EV Lac 	&	 2004-Set 2 	&	 53265.2587(0.0011) 	&	 10.315(0.013) 	&	 1.137(0.086) 	&	 1.533(0.031) 	&	 1.164(0.011) \\
EV Lac 	&	 2004-Set 2 	&	 53280.2741(0.0049) 	&	 10.306(0.015) 	&	 1.259(0.046) 	&	 1.564(0.006) 	&	 1.152(0.014) \\
EV Lac 	&	 2004-Set 2 	&	 53282.4419(0.0058) 	&	 10.306(0.011) 	&	 1.240(0.062) 	&	 1.571(0.016) 	&	 1.155(0.008) \\
EV Lac 	&	 2004-Set 2 	&	 53287.3673(0.0047) 	&	 10.311(0.009) 	&	 1.184(0.053) 	&	 1.549(0.011) 	&	 1.157(0.009) \\
EV Lac 	&	 2004-Set 2 	&	 53288.2904(0.0043) 	&	 10.328(0.009) 	&	 1.355(0.027) 	&	 1.582(0.013) 	&	 1.161(0.011) \\
EV Lac 	&	 2004-Set 2 	&	 53289.2385(0.0047) 	&	 10.291(0.008) 	&	 1.176(0.052) 	&	 1.559(0.014) 	&	 1.157(0.010) \\
EV Lac 	&	 2004-Set 2 	&	 53302.2839(0.0074) 	&	 10.304(0.012) 	&	 1.195(0.067) 	&	 1.550(0.012) 	&	 1.160(0.007) \\
EV Lac 	&	 2004-Set 2 	&	 53303.2428(0.0047) 	&	 10.284(0.009) 	&	 1.226(0.076) 	&	 1.549(0.008) 	&	 1.166(0.010) \\
EV Lac 	&	 2004-Set 2 	&	 53304.3083(0.0076) 	&	 10.288(0.011) 	&	 0.914(0.086) 	&	 1.557(0.019) 	&	 1.162(0.010) \\
EV Lac 	&	 2004-Set 2 	&	 53309.2371(0.0046) 	&	 10.328(0.008) 	&	 1.341(0.089) 	&	 1.563(0.008) 	&	 1.164(0.008) \\
EV Lac 	&	 2004-Set 2 	&	 53310.2340(0.0043) 	&	 10.327(0.007) 	&	 1.305(0.042) 	&	 1.561(0.017) 	&	 1.156(0.008) \\
EV Lac 	&	 2004-Set 2 	&	 53312.2830(0.0047) 	&	 10.288(0.009) 	&	 1.103(0.076) 	&	 1.551(0.016) 	&	 1.155(0.011) \\
EV Lac 	&	 2005-Set 3 	&	 53554.4680(0.0045) 	&	 10.302(0.011) 	&	 1.133(0.088) 	&	 1.542(0.012) 	&	 1.149(0.005) \\
EV Lac 	&	 2005-Set 3 	&	 53557.4785(0.0053) 	&	 10.330(0.008) 	&	 1.198(0.048) 	&	 1.549(0.009) 	&	 1.163(0.007) \\
EV Lac 	&	 2005-Set 3 	&	 53564.4841(0.0040) 	&	 10.290(0.006) 	&	 1.229(0.064) 	&	 1.558(0.007) 	&	 1.141(0.005) \\
EV Lac 	&	 2005-Set 3 	&	 53566.3986(0.0051) 	&	 10.330(0.008) 	&	 1.227(0.055) 	&	 1.561(0.008) 	&	 1.160(0.017) \\
EV Lac 	&	 2005-Set 3 	&	 53578.3593(0.0042) 	&	 10.312(0.009) 	&	 1.185(0.045) 	&	 1.567(0.007) 	&	 1.137(0.005) \\
EV Lac 	&	 2005-Set 3 	&	 53581.5294(0.0038) 	&	 10.310(0.008) 	&	 1.085(0.014) 	&	 1.525(0.007) 	&	 1.163(0.006) \\
EV Lac 	&	 2005-Set 3 	&	 53584.4970(0.0040) 	&	 10.315(0.006) 	&	 1.207(0.015) 	&	 1.556(0.009) 	&	 1.154(0.005) \\
EV Lac 	&	 2005-Set 3 	&	 53585.5324(0.0036) 	&	 10.297(0.010) 	&	 1.293(0.074) 	&	 1.545(0.011) 	&	 1.159(0.009) \\
EV Lac 	&	 2005-Set 3 	&	 53586.3264(0.0008) 	&	 10.300(0.004) 	&	 1.062(0.016) 	&	 1.549(0.003) 	&	 1.133(0.010) \\
EV Lac 	&	 2005-Set 3 	&	 53597.4090(0.0017) 	&	 10.304(0.008) 	&	 1.259(0.050) 	&	 1.564(0.011) 	&	 1.153(0.009) \\
EV Lac 	&	 2005-Set 3 	&	 53602.4255(0.0044) 	&	 10.314(0.006) 	&	 0.532(0.042) 	&	 1.540(0.011) 	&	 1.167(0.014) \\
EV Lac 	&	 2005-Set 3 	&	 53605.3681(0.0038) 	&	 10.337(0.011) 	&	 1.217(0.018) 	&	 1.543(0.019) 	&	 1.171(0.007) \\
EV Lac 	&	 2005-Set 3 	&	 53606.2978(0.0035) 	&	 10.313(0.007) 	&	 1.183(0.029) 	&	 1.553(0.007) 	&	 1.151(0.008) \\
EV Lac 	&	 2005-Set 3 	&	 53621.3298(0.0051) 	&	 10.296(0.007) 	&	 0.859(0.030) 	&	 1.555(0.006) 	&	 1.152(0.007) \\
EV Lac 	&	 2005-Set 3 	&	 53626.2689(0.0037) 	&	 10.328(0.001) 	&	 1.161(0.046) 	&	 1.549(0.003) 	&	 1.164(0.002) \\
EV Lac 	&	 2005-Set 3 	&	 53641.2634(0.0040) 	&	 10.297(0.005) 	&	 0.900(0.013) 	&	 1.579(0.006) 	&	 1.158(0.011) \\
EV Lac 	&	 2005-Set 3 	&	 53647.2519(0.0038) 	&	 10.303(0.007) 	&	 1.138(0.026) 	&	 1.540(0.010) 	&	 1.161(0.013) \\
EV Lac 	&	 2005-Set 3 	&	 53652.2573(0.0037) 	&	 10.313(0.009) 	&	 1.092(0.042) 	&	 1.549(0.009) 	&	 1.146(0.010) \\
EV Lac 	&	 2006-Set 4 	&	 53863.5667(0.0029) 	&	 10.274(0.004) 	&	 1.244(0.014) 	&	 1.567(0.010) 	&	 1.154(0.006) \\
EV Lac 	&	 2006-Set 4 	&	 53877.5499(0.0027) 	&	 10.285(0.014) 	&	 1.092(0.078) 	&	 1.571(0.012) 	&	 1.152(0.012) \\
EV Lac 	&	 2006-Set 4 	&	 53883.5378(0.0038) 	&	 10.372(0.006) 	&	 1.069(0.062) 	&	 1.582(0.007) 	&	 1.158(0.006) \\
EV Lac 	&	 2006-Set 4 	&	 53886.4627(0.0034) 	&	 10.295(0.006) 	&	 1.130(0.040) 	&	 1.542(0.008) 	&	 1.167(0.005) \\
EV Lac 	&	 2006-Set 4 	&	 53894.5260(0.0028) 	&	 10.267(0.004) 	&	 1.215(0.060) 	&	 1.559(0.005) 	&	 1.147(0.004) \\
EV Lac 	&	 2006-Set 4 	&	 53907.5303(0.0036) 	&	 10.264(0.005) 	&	 1.163(0.034) 	&	 1.571(0.009) 	&	 1.128(0.008) \\
EV Lac 	&	 2006-Set 4 	&	 53908.5431(0.0031) 	&	 10.299(0.002) 	&	 1.176(0.015) 	&	 1.566(0.004) 	&	 1.146(0.003) \\
EV Lac 	&	 2006-Set 4 	&	 53915.5406(0.0030) 	&	 10.306(0.004) 	&	 1.125(0.033) 	&	 1.538(0.006) 	&	 1.152(0.005) \\
EV Lac 	&	 2006-Set 4 	&	 53938.5340(0.0038) 	&	 10.277(0.006) 	&	 1.204(0.048) 	&	 1.559(0.006) 	&	 1.145(0.007) \\
EV Lac 	&	 2006-Set 4 	&	 53940.4593(0.0038) 	&	 10.382(0.003) 	&	 1.121(0.073) 	&	 1.580(0.008) 	&	 1.172(0.004) \\
EV Lac 	&	 2006-Set 4 	&	 53946.4700(0.0008) 	&	 10.260(0.002) 	&	 1.030(0.063) 	&	 1.550(0.002) 	&	 1.135(0.005) \\
EV Lac 	&	 2006-Set 4 	&	 53950.4652(0.0008) 	&	 10.303(0.005) 	&	 1.231(0.043) 	&	 1.545(0.008) 	&	 1.139(0.007) \\
EV Lac 	&	 2006-Set 4 	&	 53951.4800(0.0026) 	&	 10.263(0.003) 	&	 1.172(0.053) 	&	 1.555(0.007) 	&	 1.133(0.005) \\
EV Lac 	&	 2006-Set 4 	&	 53955.3607(0.0039) 	&	 10.253(0.007) 	&	 1.258(0.087) 	&	 1.579(0.007) 	&	 1.135(0.007) \\
EV Lac 	&	 2006-Set 4 	&	 53956.3324(0.0034) 	&	 10.274(0.011) 	&	 0.901(0.033) 	&	 1.581(0.012) 	&	 1.137(0.009) \\
EV Lac 	&	 2006-Set 4 	&	 53961.4980(0.0027) 	&	 10.321(0.004) 	&	 1.117(0.011) 	&	 1.571(0.007) 	&	 1.148(0.004) \\
EV Lac 	&	 2006-Set 4 	&	 53963.3328(0.0032) 	&	 10.322(0.003) 	&	 0.931(0.036) 	&	 1.546(0.008) 	&	 1.160(0.006) \\
EV Lac 	&	 2006-Set 4 	&	 53972.3414(0.0035) 	&	 10.292(0.006) 	&	 1.199(0.012) 	&	 1.567(0.006) 	&	 1.144(0.011) \\
EV Lac 	&	 2006-Set 4 	&	 53973.4465(0.0034) 	&	 10.268(0.003) 	&	 1.287(0.022) 	&	 1.557(0.005) 	&	 1.149(0.004) \\
EV Lac 	&	 2006-Set 4 	&	 53984.2589(0.0008) 	&	 10.390(0.007) 	&	 1.146(0.083) 	&	 1.562(0.008) 	&	 1.179(0.011) \\
EV Lac 	&	 2006-Set 4 	&	 53988.2519(0.0027) 	&	 10.360(0.003) 	&	 1.009(0.044) 	&	 1.573(0.007) 	&	 1.172(0.007) \\
EV Lac 	&	 2006-Set 4 	&	 53992.4226(0.0027) 	&	 10.358(0.003) 	&	 1.074(0.036) 	&	 1.550(0.003) 	&	 1.160(0.006) \\
EV Lac 	&	 2006-Set 4 	&	 54009.2659(0.0027) 	&	 10.314(0.003) 	&	 1.158(0.055) 	&	 1.571(0.013) 	&	 1.157(0.003) \\
EV Lac 	&	 2006-Set 4 	&	 54037.2116(0.0037) 	&	 10.388(0.005) 	&	 1.145(0.056) 	&	 1.565(0.012) 	&	 1.150(0.009) \\
EV Lac 	&	 2006-Set 4 	&	 54047.2046(0.0027) 	&	 10.271(0.008) 	&	 1.112(0.026) 	&	 1.541(0.014) 	&	 1.142(0.010) \\
EV Lac 	&	 2006-Set 4 	&	 54055.2251(0.0027) 	&	 10.326(0.008) 	&	 1.159(0.043) 	&	 1.553(0.018) 	&	 1.152(0.012) \\
EV Lac 	&	 2006-Set 4 	&	 54058.2051(0.0000) 	&	 10.364(0.002) 	&	 1.146(0.088) 	&	 1.587(0.007) 	&	 1.145(0.007) \\
\hline
V1005 Ori 	&	 2004/2005 	&	 53353.4006(0.0037) 	&	 10.184(0.011) 	&	 1.152(0.053) 	&	 1.306(0.010) 	&	 0.882(0.008) \\
V1005 Ori 	&	 2004/2005 	&	 53355.3312(0.0047) 	&	 10.253(0.012) 	&	 1.232(0.048) 	&	 1.313(0.014) 	&	 0.892(0.011) \\
V1005 Ori 	&	 2004/2005 	&	 53374.3572(0.0030) 	&	 10.125(0.007) 	&	 1.151(0.044) 	&	 1.316(0.009) 	&	 0.877(0.008) \\
V1005 Ori 	&	 2004/2005 	&	 53376.3710(0.0035) 	&	 10.206(0.010) 	&	 1.260(0.057) 	&	 1.314(0.016) 	&	 0.886(0.012) \\
V1005 Ori 	&	 2004/2005 	&	 53377.2512(0.0000) 	&	 10.246(0.010) 	&	 1.207(0.038) 	&	 1.319(0.017) 	&	 0.897(0.012) \\
V1005 Ori 	&	 2004/2005 	&	 53381.3452(0.0038) 	&	 10.256(0.007) 	&	 1.252(0.052) 	&	 1.313(0.016) 	&	 0.903(0.011) \\
V1005 Ori 	&	 2004/2005 	&	 53383.2424(0.0045) 	&	 10.116(0.006) 	&	 0.997(0.031) 	&	 1.293(0.010) 	&	 0.877(0.008) \\
V1005 Ori 	&	 2004/2005 	&	 53412.2611(0.0040) 	&	 10.274(0.006) 	&	 1.212(0.015) 	&	 1.319(0.005) 	&	 0.917(0.014) \\
V1005 Ori 	&	 2004/2005 	&	 53413.2330(0.0042) 	&	 10.234(0.004) 	&	 1.257(0.075) 	&	 1.315(0.008) 	&	 0.897(0.009) \\
V1005 Ori 	&	 2004/2005 	&	 53414.2662(0.0007) 	&	 10.176(0.007) 	&	 1.258(0.042) 	&	 1.317(0.014) 	&	 0.898(0.006) \\
V1005 Ori 	&	 2004/2005 	&	 53453.2770(0.0046) 	&	 10.189(0.008) 	&	 0.830(0.051) 	&	 1.331(0.008) 	&	 0.894(0.009) \\
V1005 Ori 	&	 2005/2006 	&	 53640.5712(0.0039) 	&	 10.211(0.005) 	&	 1.168(0.046) 	&	 1.325(0.008) 	&	 0.884(0.006) \\
V1005 Ori 	&	 2005/2006 	&	 53647.5820(0.0038) 	&	 10.188(0.006) 	&	 1.232(0.089) 	&	 1.319(0.007) 	&	 0.880(0.007) \\
V1005 Ori 	&	 2005/2006 	&	 53664.5827(0.0062) 	&	 10.181(0.012) 	&	 1.179(0.061) 	&	 1.306(0.033) 	&	 0.883(0.006) \\
V1005 Ori 	&	 2005/2006 	&	 53669.4940(0.0046) 	&	 10.185(0.004) 	&	 1.100(0.052) 	&	 1.320(0.010) 	&	 0.878(0.005) \\
V1005 Ori 	&	 2005/2006 	&	 53672.4511(0.0039) 	&	 10.181(0.008) 	&	 1.162(0.034) 	&	 1.303(0.008) 	&	 0.878(0.006) \\
V1005 Ori 	&	 2005/2006 	&	 53673.4626(0.0036) 	&	 10.191(0.013) 	&	 1.132(0.014) 	&	 1.329(0.030) 	&	 0.872(0.011) \\
V1005 Ori 	&	 2005/2006 	&	 53686.5181(0.0040) 	&	 10.202(0.003) 	&	 1.310(0.025) 	&	 1.339(0.007) 	&	 0.892(0.007) \\
V1005 Ori 	&	 2005/2006 	&	 53702.3409(0.0042) 	&	 10.216(0.006) 	&	 1.173(0.052) 	&	 1.317(0.014) 	&	 0.896(0.002) \\
V1005 Ori 	&	 2005/2006 	&	 53717.3477(0.0065) 	&	 10.213(0.007) 	&	 1.150(0.025) 	&	 1.338(0.016) 	&	 0.910(0.013) \\
V1005 Ori 	&	 2005/2006 	&	 53724.3014(0.0040) 	&	 10.211(0.002) 	&	 1.259(0.045) 	&	 1.305(0.005) 	&	 0.875(0.004) \\
V1005 Ori 	&	 2005/2006 	&	 53725.2527(0.0035) 	&	 10.194(0.005) 	&	 1.226(0.017) 	&	 1.298(0.003) 	&	 0.874(0.005) \\
V1005 Ori 	&	 2005/2006 	&	 53729.2900(0.0036) 	&	 10.207(0.002) 	&	 1.209(0.055) 	&	 1.306(0.006) 	&	 0.892(0.003) \\
V1005 Ori 	&	 2005/2006 	&	 53737.2813(0.0046) 	&	 10.224(0.004) 	&	 1.253(0.021) 	&	 1.301(0.002) 	&	 0.893(0.004) \\
V1005 Ori 	&	 2005/2006 	&	 53744.3067(0.0068) 	&	 10.213(0.005) 	&	 1.361(0.045) 	&	 1.340(0.004) 	&	 0.906(0.004) \\
V1005 Ori 	&	 2005/2006 	&	 53757.3701(0.0028) 	&	 10.184(0.002) 	&	 1.313(0.018) 	&	 1.308(0.003) 	&	 0.884(0.002) \\
V1005 Ori 	&	 2005/2006 	&	 53763.2739(0.0042) 	&	 10.225(0.004) 	&	 1.143(0.013) 	&	 1.330(0.005) 	&	 0.901(0.005) \\
V1005 Ori 	&	 2005/2006 	&	 53764.2990(0.0036) 	&	 10.216(0.003) 	&	 1.202(0.025) 	&	 1.317(0.004) 	&	 0.887(0.003) \\
V1005 Ori 	&	 2005/2006 	&	 53765.4071(0.0030) 	&	 10.207(0.003) 	&	 1.195(0.008) 	&	 1.302(0.007) 	&	 0.907(0.004) \\
V1005 Ori 	&	 2005/2006 	&	 53771.3851(0.0039) 	&	 10.178(0.004) 	&	 1.256(0.035) 	&	 1.312(0.008) 	&	 0.883(0.002) \\
V1005 Ori 	&	 2005/2006 	&	 53788.2932(0.0030) 	&	 10.160(0.003) 	&	 1.217(0.035) 	&	 1.328(0.006) 	&	 0.877(0.004) \\
V1005 Ori 	&	 2005/2006 	&	 53796.2760(0.0035) 	&	 10.197(0.005) 	&	 1.177(0.048) 	&	 1.309(0.011) 	&	 0.887(0.006) \\
V1005 Ori 	&	 2005/2006 	&	 53812.2458(0.0030) 	&	 10.253(0.005) 	&	 1.231(0.052) 	&	 1.302(0.006) 	&	 0.906(0.003) \\
V1005 Ori 	&	 2006/2007 	&	 53984.5749(0.0037) 	&	 10.237(0.005) 	&	 1.161(0.046) 	&	 1.312(0.008) 	&	 0.868(0.006) \\
V1005 Ori 	&	 2006/2007 	&	 53987.5873(0.0035) 	&	 10.185(0.008) 	&	 1.341(0.049) 	&	 1.312(0.005) 	&	 0.880(0.008) \\
V1005 Ori 	&	 2006/2007 	&	 53988.5834(0.0038) 	&	 10.250(0.011) 	&	 1.341(0.010) 	&	 1.331(0.011) 	&	 0.884(0.013) \\
V1005 Ori 	&	 2006/2007 	&	 53994.5802(0.0037) 	&	 10.180(0.006) 	&	 1.237(0.070) 	&	 1.304(0.009) 	&	 0.866(0.008) \\
V1005 Ori 	&	 2006/2007 	&	 54033.5946(0.0038) 	&	 10.259(0.004) 	&	 1.146(0.042) 	&	 1.313(0.009) 	&	 0.895(0.006) \\
V1005 Ori 	&	 2006/2007 	&	 54047.4878(0.0038) 	&	 10.183(0.010) 	&	 1.154(0.084) 	&	 1.301(0.026) 	&	 0.886(0.008) \\
V1005 Ori 	&	 2006/2007 	&	 54049.5410(0.0035) 	&	 10.184(0.011) 	&	 1.110(0.029) 	&	 1.304(0.012) 	&	 0.871(0.016) \\
V1005 Ori 	&	 2006/2007 	&	 54064.5743(0.0035) 	&	 10.265(0.010) 	&	 1.225(0.057) 	&	 1.319(0.010) 	&	 0.890(0.009) \\
V1005 Ori 	&	 2006/2007 	&	 54066.3884(0.0035) 	&	 10.145(0.017) 	&	 1.266(0.070) 	&	 1.304(0.009) 	&	 0.882(0.010) \\
V1005 Ori 	&	 2006/2007 	&	 54071.3212(0.0035) 	&	 10.155(0.010) 	&	 1.267(0.049) 	&	 1.337(0.010) 	&	 0.868(0.010) \\
V1005 Ori 	&	 2006/2007 	&	 54085.5616(0.0049) 	&	 10.225(0.013) 	&	 1.289(0.037) 	&	 1.331(0.016) 	&	 0.868(0.013) \\
V1005 Ori 	&	 2006/2007 	&	 54095.4691(0.0037) 	&	 10.273(0.007) 	&	 1.237(0.017) 	&	 1.315(0.007) 	&	 0.899(0.006) \\
V1005 Ori 	&	 2006/2007 	&	 54109.3354(0.0035) 	&	 10.212(0.006) 	&	 1.208(0.078) 	&	 1.297(0.007) 	&	 0.893(0.006) \\
V1005 Ori 	&	 2006/2007 	&	 54122.3280(0.0035) 	&	 10.225(0.005) 	&	 1.221(0.077) 	&	 1.312(0.005) 	&	 0.884(0.010) \\
V1005 Ori 	&	 2006/2007 	&	 54158.2553(0.0038) 	&	 10.140(0.013) 	&	 1.127(0.023) 	&	 1.315(0.018) 	&	 0.863(0.014) \\
\hline
V1054 Oph	&	2004	&	53136.5478(0.0013)	&	8.996(0.012)	&	1.147(0.005)	&	1.548(0.007)	&	\\
V1054 Oph	&	2004	&	53138.4673(0.0010)	&	8.991(0.012)	&	1.141(0.005)	&	1.542(0.006)	&	\\
V1054 Oph	&	2004	&	53146.5107(0.0012)	&	8.989(0.008)	&	1.149(0.005)	&	1.545(0.006)	&	\\
V1054 Oph	&	2004	&	53151.4962(0.0017)	&	8.996(0.012)	&	1.161(0.006)	&	1.543(0.006)	&	\\
V1054 Oph	&	2004	&	53152.4609(0.0013)	&	8.996(0.011)	&	1.161(0.006)	&	1.548(0.006)	&	\\
V1054 Oph	&	2004	&	53153.5004(0.0017)	&	8.989(0.008)	&	1.144(0.006)	&	1.545(0.007)	&	\\
V1054 Oph	&	2004	&	53157.5139(0.0020)	&	8.988(0.010)	&	1.138(0.006)	&	1.547(0.007)	&	\\
V1054 Oph	&	2004	&	53163.5248(0.0015)	&	8.995(0.008)	&	1.142(0.006)	&	1.545(0.006)	&	\\
V1054 Oph	&	2004	&	53167.4368(0.0015)	&	8.991(0.008)	&	1.146(0.005)	&	1.542(0.005)	&	\\
V1054 Oph	&	2005	&	53505.4608(0.0004)	&	9.004(0.004)	&	1.066(0.006)	&	1.555(0.005)	&	1.062(0.004)\\
V1054 Oph	&	2005	&	53526.3480(0.0005)	&	9.000(0.008)	&	1.074(0.006)	&	1.548(0.005)	&	1.064(0.004)\\
V1054 Oph	&	2005	&	53527.3366(0.0004)	&	8.998(0.009)	&	1.055(0.006)	&	1.545(0.005)	&	1.064(0.004)\\
V1054 Oph	&	2005	&	53528.3373(0.0004)	&	9.004(0.007)	&	1.111(0.006)	&	1.548(0.005)	&	1.061(0.005)\\
V1054 Oph	&	2005	&	53546.3539(0.0004)	&	9.000(0.005)	&	1.120(0.006)	&	1.551(0.005)	&	1.060(0.004)\\
V1054 Oph	&	2005	&	53547.3576(0.0004)	&	9.006(0.008)	&	1.092(0.007)	&	1.555(0.006)	&	1.060(0.005)\\
V1054 Oph	&	2005	&	53548.3520(0.0000)	&	8.996(0.009)	&	1.070(0.006)	&	1.554(0.005)	&	1.058(0.005)\\
V1054 Oph	&	2005	&	53557.3294(0.0004)	&	9.003(0.010)	&	1.056(0.006)	&	1.545(0.006)	&	1.066(0.005)\\
V1054 Oph	&	2005	&	53564.3437(0.0018)	&	8.998(0.007)	&	1.079(0.005)	&	1.550(0.004)	&	1.060(0.004)\\
V1054 Oph	&	2006	&	53861.4662(0.0003)	&	8.994(0.008)	&	1.134(0.006)	&	1.543(0.005)	&	1.064(0.004)\\
V1054 Oph	&	2006	&	53863.3951(0.0004)	&	8.989(0.007)	&	1.145(0.006)	&	1.556(0.005)	&	1.064(0.005)\\
V1054 Oph	&	2006	&	53875.3481(0.0004)	&	8.997(0.006)	&	1.165(0.005)	&	1.551(0.004)	&	1.065(0.004)\\
V1054 Oph	&	2006	&	53877.3435(0.0004)	&	8.994(0.006)	&	1.143(0.006)	&	1.542(0.005)	&	1.057(0.005)\\
V1054 Oph	&	2006	&	53882.3311(0.0004)	&	8.986(0.005)	&	1.187(0.005)	&	1.556(0.004)	&	1.061(0.004)\\
V1054 Oph	&	2006	&	53893.3528(0.0003)	&	8.995(0.005)	&	1.168(0.005)	&	1.554(0.004)	&	1.059(0.004)\\
V1054 Oph	&	2006	&	53894.3664(0.0004)	&	8.991(0.004)	&	1.127(0.005)	&	1.549(0.005)	&	1.061(0.005)\\
\hline
EQ Peg	&	2004	&	53236.3967(0.0011)	&	10.157(0.008)	&	0.963(0.014)	&	1.574(0.013)	&	1.183(0.012)\\
EQ Peg	&	2004	&	53237.3662(0.0011)	&	10.129(0.007)	&	0.933(0.016)	&	1.598(0.012)	&	1.130(0.013)\\
EQ Peg	&	2004	&	53259.4116(0.0011)	&	10.149(0.007)	&	0.939(0.011)	&	1.568(0.011)	&	1.166(0.016)\\
EQ Peg	&	2004	&	53260.4846(0.0008)	&	10.173(0.006)	&	1.053(0.015)	&	1.560(0.016)	&	1.179(0.013)\\
EQ Peg	&	2004	&	53261.6059(0.0008)	&	10.144(0.007)	&	0.947(0.011)	&	1.576(0.011)	&	1.160(0.013)\\
EQ Peg	&	2004	&	53263.3286(0.0008)	&	10.165(0.008)	&	1.011(0.010)	&	1.597(0.014)	&	1.178(0.013)\\
EQ Peg	&	2004	&	53264.3962(0.0023)	&	10.186(0.008)	&	0.931(0.016)	&	1.578(0.016)	&	1.165(0.013)\\
EQ Peg	&	2004	&	53265.3348(0.0007)	&	10.197(0.006)	&	1.079(0.016)	&	1.567(0.013)	&	1.172(0.016)\\
EQ Peg	&	2004	&	53280.3039(0.0050)	&	10.162(0.006)	&	1.052(0.015)	&	1.587(0.011)	&	1.174(0.012)\\
EQ Peg	&	2004	&	53281.2675(0.0022)	&	10.180(0.008)	&	1.107(0.017)	&	1.591(0.010)	&	1.160(0.013)\\
EQ Peg	&	2004	&	53287.3928(0.0022)	&	10.142(0.008)	&	1.070(0.011)	&	1.581(0.013)	&	1.157(0.013)\\
EQ Peg	&	2004	&	53289.3472(0.0047)	&	10.154(0.006)	&	1.125(0.013)	&	1.574(0.012)	&	1.163(0.012)\\
EQ Peg	&	2004	&	53335.2905(0.0010)	&	10.169(0.008)	&	0.864(0.012)	&	1.565(0.014)	&	1.148(0.013)\\
EQ Peg	&	2005	&	53621.3608(0.0030)	&	10.144(0.006)	&	1.021(0.015)	&	1.561(0.013)	&	1.155(0.012)\\
EQ Peg	&	2005	&	53622.4013(0.0029)	&	10.146(0.005)	&	0.992(0.017)	&	1.567(0.011)	&	1.150(0.011)\\
EQ Peg	&	2005	&	53626.2942(0.0012)	&	10.160(0.005)	&	1.127(0.013)	&	1.562(0.012)	&	1.138(0.012)\\
EQ Peg	&	2005	&	53641.2928(0.0030)	&	10.163(0.005)	&	1.127(0.012)	&	1.554(0.011)	&	1.158(0.011)\\
EQ Peg	&	2005	&	53652.2815(0.0027)	&	10.144(0.007)	&	1.138(0.015)	&	1.546(0.012)	&	1.151(0.011)\\
EQ Peg	&	2005	&	53664.2628(0.0034)	&	10.175(0.005)	&	1.002(0.013)	&	1.556(0.012)	&	1.175(0.011)\\
EQ Peg	&	2005	&	53672.2250(0.0043)	&	10.153(0.006)	&	1.094(0.014)	&	1.571(0.012)	&	1.137(0.012)\\
EQ Peg	&	2005	&	53673.2791(0.0030)	&	10.158(0.007)	&	1.095(0.014)	&	1.585(0.020)	&	1.157(0.013)\\
EQ Peg	&	2005	&	53686.2577(0.0028)	&	10.164(0.006)	&	1.067(0.015)	&	1.560(0.012)	&	1.144(0.011)\\
\hline
\end{longtable}

The time series analyses do not demonstrate any short-term variation in one season for V1054 Oph. However, the levels of both the brightness in V band and the colours are varying from one season to next one. This is seen from Figure 1. The light and colour curves in this figure are versus HJD instead of phase. This is because there is no known rotational period for V1054 Oph. As it is seen from the figure, there is no data for V-R colour variation in the first observing season of the star, this is because the star was observed in the UBV bands in the first season. Like V1054 Oph, the time series analyses do not reveal any regular variation for AD Leo. For other analyses, the ephemeris given in Equation (1) taken from \citet{Pan93} was used in phase calculations for all UBVR observations of AD Leo.

\begin{center}
\begin{equation}
JD~(Hel.)~=~24~49099.489~+~2^{d}.7~\times~E.
\end{equation}
\end{center}
In Figure 2, standard V band light and U-B, B-V and V-R colour curves are shown for three observing seasons. When this Figure is examined, it is clearly seen that some observation points exceed above the level of the standard deviations, but the time series analysis does not give any regular variation. Although AD Leo does not show any regular variation in one observing season, as it is seen from the Figure, level of the brightness in V band is increasing about $0^{m}.01$ from the season 2004/2005 to next one and $0^{m}.02$ from the season 2005/2006 to last season. This is the same in the other bands. The colour curves in Figure 2 show that the stars do no exhibit any distinctive colour variation in a season. On the other hand, V-R colour gets bluer from a season to next one, when the star get brighter.

\begin{figure*}
\hspace{15mm}
\vspace{5mm}
\FigureFile(155mm,60mm){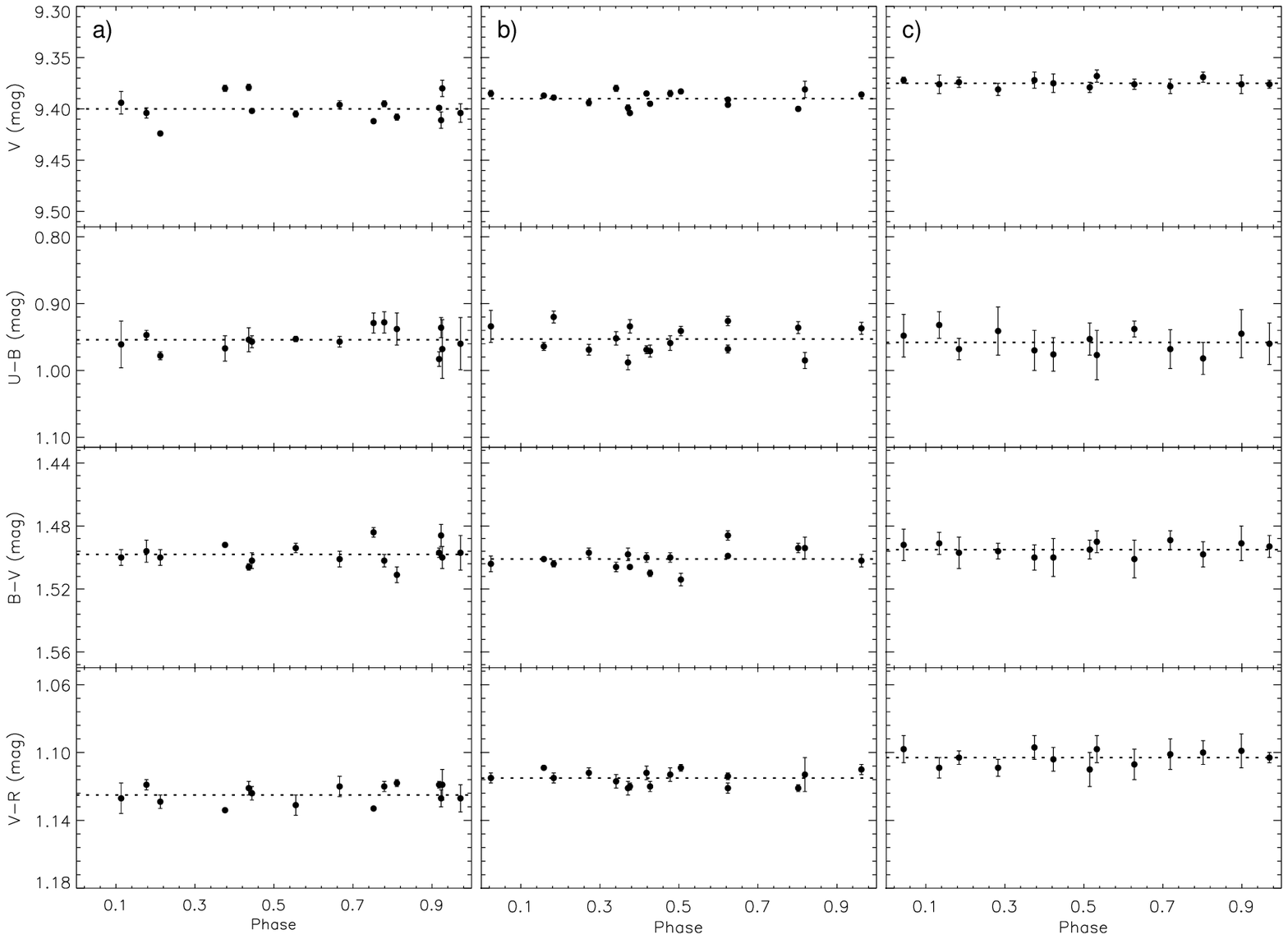}
\caption{The light and colour curves obtained from the observations of AD Leo in this study are seen for the seasons 2004/2005 (a), 2005/2006 (b) and 2006/2007 (c). Dashed lines represent the mean brightness and colour level obtained from the time series analyses.\label{fig2}}
\end{figure*}

In case of EV Lac, the time series analyses gave almost the same periods changing unsystematically from $4^{d}.331\pm0.037$ to $4^{d}.379\pm0.017$ for each data set of three observing season. When all the data sets were analysed together, the time series analysis gave a period of $4^{d}.3517\pm0.001$ given in Equation (2) for rotational modulation of EV Lac. The periods found from the each sets and the all seasons are similar to those found by \citet{Pet83} and \citet{Mah81}.

It must be noted that we could not observed EV Lac for a while, about 20 days, in the middle of the seasons 2004. When we started to observe the star after 20 days later, we saw that the light curves had been partly changed. Therefore, the data set was divided in two parts as Set 1 and Set 2 to solve this problem. We think that this is because of the rapid variations of active areas on the stellar surface of EV Lac. We did not see any similar variations on the other program stars. For example, we could not observed AD Leo for almost 100 days, in the middle of some seasons, but we saw that the light curves had not been changed yet. This is why we did not divided the data sets of AD Leo into two or more parts. This is the same for other stars.

\begin{figure*}
\hspace{15mm}
\vspace{5mm}
\FigureFile(155mm,60mm){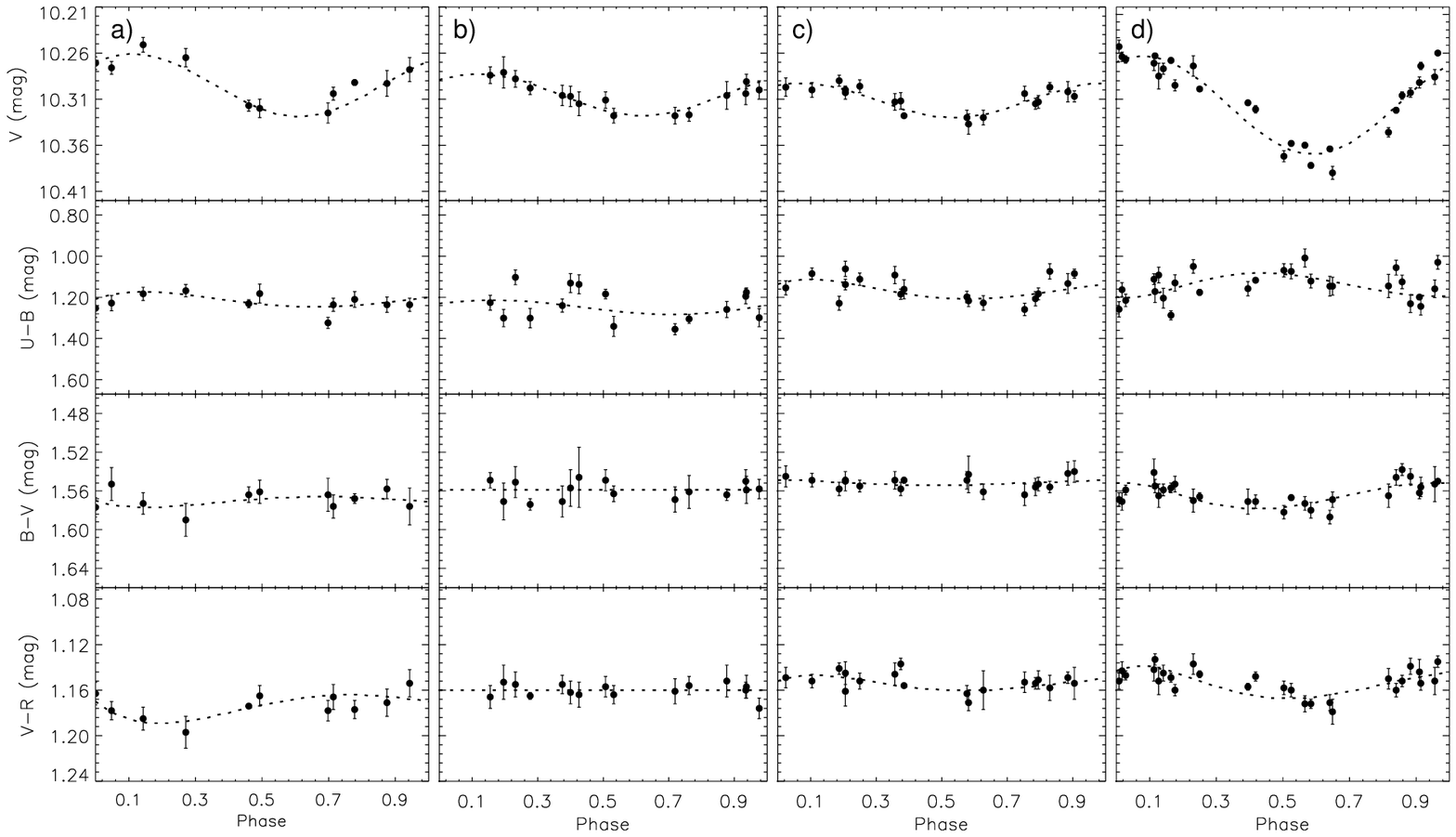}
\caption{The light and colour curves obtained in this study are seen for four data sets composed from the observations of EV Lac. a) Observing season 2004 - the first part, b) Observing season 2004 - the second part, c) Observing season 2005, d) Observing season 2006. Dashed lines represent the models derived from the Discrete Fourier Transform.\label{fig3}}
\end{figure*}

\begin{center}
\begin{equation}
JD~(Hel.)~=~24~53202.47895~+~4^{d}.3517~\times~E.
\end{equation}
\end{center}
Using the ephemeris given in Equation (2), the phases were computed for all data sets. V band light and U-B, B-V and V-R colour curves are shown for four data sets in Figure 3. For V band of EV Lac, the variations of minimum, maximum and mean average of brightness, amplitude and periods found from each data set are given in Figure 4. The time series analyses also show that minimum phases of the curves are changing. It is at about $0^{P}.62$ for Set 1 and 2, $0^{P}.54$ for Set 3, and $0^{P}.60$ for Set 4. Comparing the colour curves with light curves for Set 1, it is seen that B-V and V-R colours are getting bluer towards the phase of minimum seen in the light curve. There is no variation in colour curves for data Set 2 and 3. On the other hand, both the colours are getting reddening towards the minimum phases of light curves in the last season.

\begin{figure*}
\hspace{50mm}
\vspace{5mm}
\FigureFile(110mm,60mm){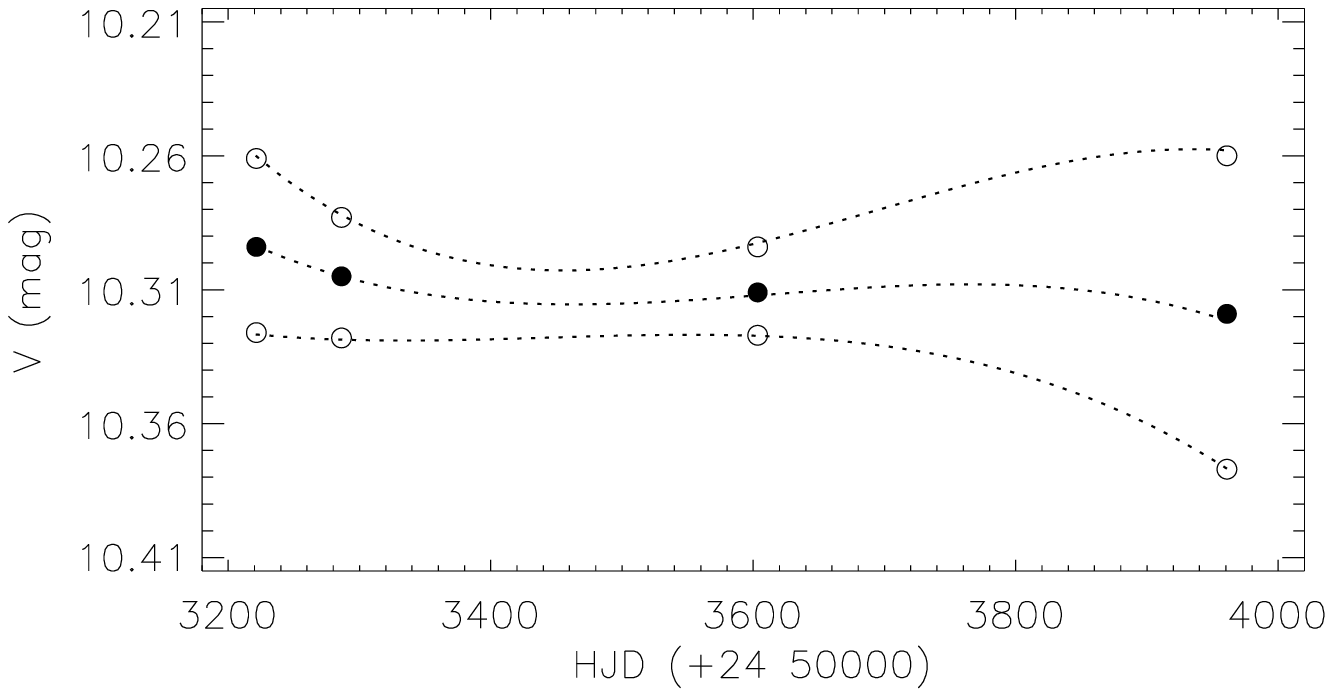}
\caption{The figure shows the results of the analyses for 4 data sets of EV Lac in V band. The variations of minimum, maximum and mean average levels of brightness are shown. In the figure, open circles show minimum and maximum levels, while filled circles show the mean average of the brightness for each data set. Dashed lines represent the fits of brightnesses to just show the variations.\label{fig4}}
\end{figure*}

V1005 Ori was observed in three observing seasons. Because of the winter weather conditions, some observation could not be carried on. This is why there are some empty phases in the light curve of the season of 2004/2005. However, the data obtained is enough for the time series analysis. The time series analyses gave almost the same periods changing unsystematically from $4^{d}.419\pm0.005$ to $4^{d}.429\pm0.014$ for each data set of three observing season. The periods found in this study are close to the period found by \citet{Byr84}. When all the data sets were analysed together, the time series analysis gave a period of $4^{d}.4236\pm0.001$ given in Equation (3) for rotational modulation of V1005 Ori.

\begin{figure*}
\hspace{15mm}
\vspace{5mm}
\FigureFile(155mm,60mm){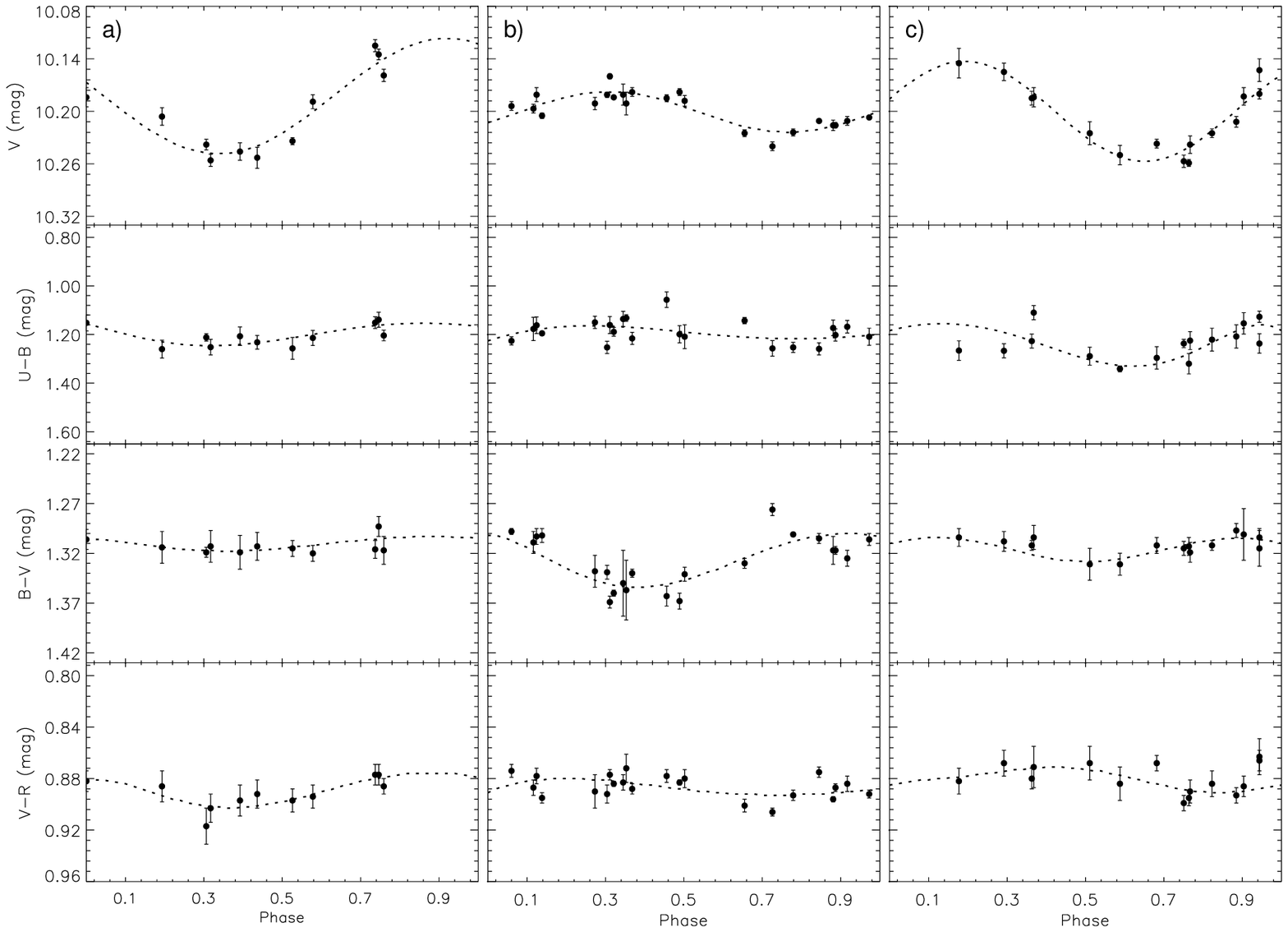}
\caption{The light and colour curves obtained from the observations of V1005 Ori in this study are seen for the seasons 2004/2005 (a), 2005/2006 (b) and 2006/2007 (c). Dashed lines represent the models derived from the Discrete Fourier Transform.\label{fig5}}
\end{figure*}

\begin{center}
\begin{equation}
JD~(Hel.)~= ~24~53353.40036~+~4^{d}.4236~\times~E.
\end{equation}
\end{center}
Using the ephemeris given in Equation (3), the phases were computed for all data sets of V1005 Ori. Standard V band light and U-B, B-V and V-R colour curves are shown for four data sets in Figure 5. For V band of V1005 Ori, the variations of minimum, maximum and mean average of brightness, amplitude and periods for each season are given in Figure 6. When the light curves of V1005 Ori are examined, the phase of minimum was about $0^{P}.34$ for the season of 2004/2005 and it was $0^{P}.66$ for the season of 2006/2007. Although the amplitude of the curve for the season of 2005/2006 was the smallest among the previous and the latter, the analyses indicate that the phase of minimum might be $0^{P}.78$. It is seen that there is no variation in the B-V colour index for the first season, while there is reddening in the V-R colour index towards the minimum phase of the light curves. On the other hand, for the season of 2005/2006, although the light curves in Johnson UBVR bands have the smallest amplitudes, an extreme excess is seen in the B-V colour. Moreover, the B-V colour is getting bluer towards the minimum phase of the light curves. The V-R colour is reddening towards the minimum phase of the light curve in this season. In the last season, both B-V and V-R colours are reddening towards the minimum phase of the light curves.

\begin{figure*}
\hspace{50mm}
\vspace{5mm}
\FigureFile(110mm,60mm){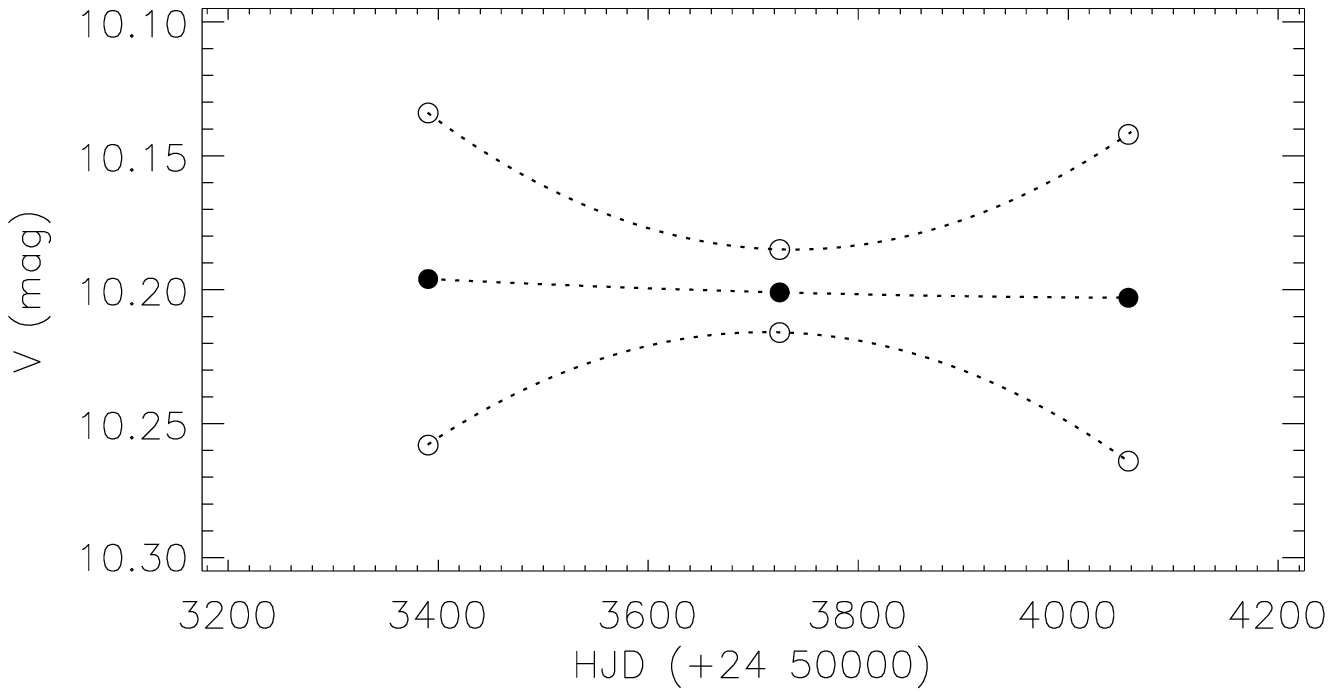}
\caption{The figure shows the results of the analyses for 3 data sets of V1005 Ori in V band. The variations of minimum, maximum and mean average levels of brightness are shown. In the figure, open circles show minimum and maximum levels, while filled circles show the mean average of the brightness for each data set. Dashed lines represent the fits of brightnesses to just show the variations.\label{fig6}}
\end{figure*}

In case of EQ Peg, the time series analyses demonstrated that EQ Peg has a variability. The analyses gave almost the same period as $1^{d}.0608\pm0.0001$ for data set combined from two observing seasons. Using the ephemeris given in Equation (4), the phases were computed for all data sets of EQ Peg.

\begin{center}
\begin{equation}
JD~(Hel.)~= ~24~53236.39669~+~1^{d}.0608~\times~E.
\end{equation}
\end{center}
Standard V band light and U-B, B-V and V-R colour curves are shown for four data sets in Figure 7. In the figure, the data sets of season 2004 and season 2005 were not separated into two panel. This is because the light curves of both seasons have the same shape. The time series analyses showed that the phases of the rotational modulation is $0^{P}.32$ for the light curves of EQ Peg. Instead of light curves, the data set combined from EQ Peg observations do not exhibits any variations.

\begin{figure*}
\hspace{65mm}
\vspace{5mm}
\FigureFile(140mm,60mm){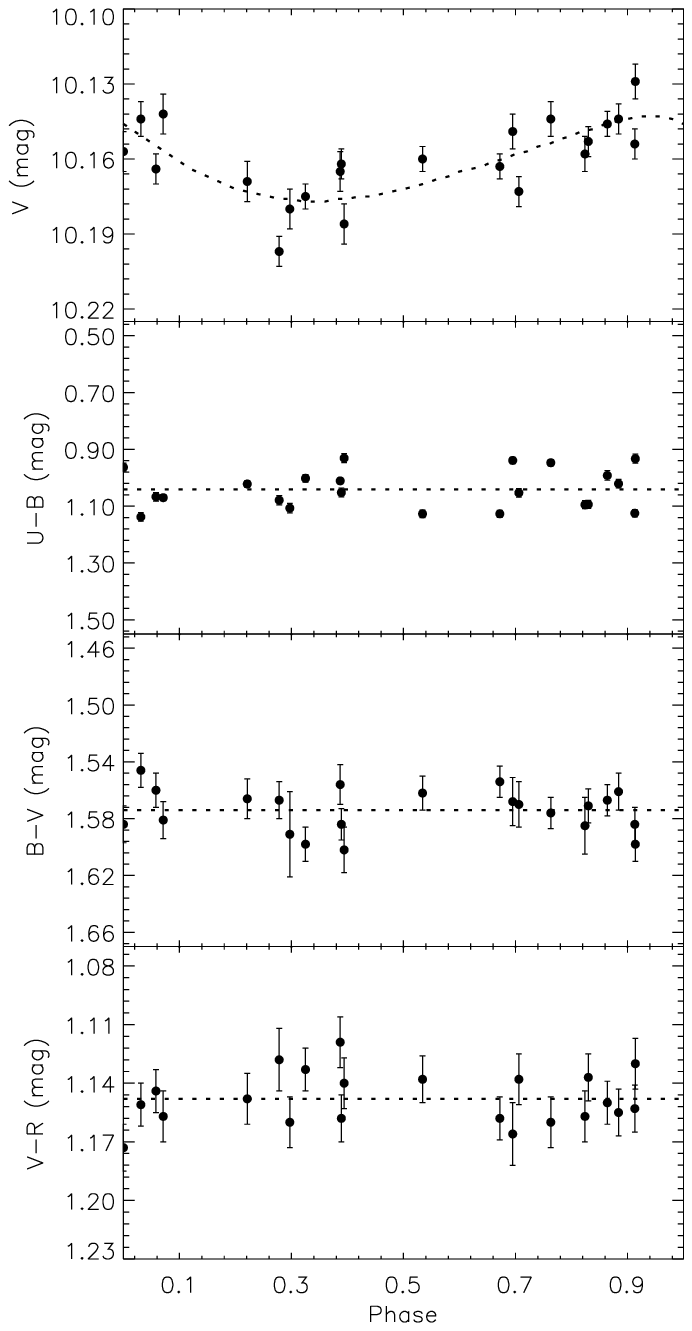}
\caption{The light and colour curves obtained from the observations of EQ Peg in this study are seen together for the seasons 2004 and 2005. Dashed lines represent the models derived from the Discrete Fourier Transform.\label{fig7}}
\end{figure*}

\subsection{The Distributions of Flare Occurrence Rates Versus Photometric Period}

These stars, for which BY Dra Syndrome is discussed, show also high level flare activity. To investigate whether there is any relation between stellar spots and flare activities, the rates of flare occurrence versus rotational phase and the phases of the minima of the V band light curves were compared for each star. V band light curves for each star were obtained in this study. The flare data of the stars have been given by \citet{Dal10} and \citet{Dal11}.

To carry out this investigation, first of all, the phases of flare maxima were computed for all flare types (together with fast and slow flares) with the same method used for the phases of light curves. The flare maximum times were used to compute the phases due to main energy emitting in this part of the flare light curves. In addition, the periods of stars are too long according to the average of flare total durations. In the second step, computing the ratio of flare number to monitoring time in intervals of 0.10 phase length as the same method used by \citet{Let97}, the mean flare occurence rate was derived for data set of each star with using Equation (5).

\begin{center}
\begin{equation}
N~=~\Sigma n_{f}~/~\Sigma T_{t}
\end{equation}
\end{center}
where $N$ is the mean flare occurence rate in intervals of 0.10 phase length. $n_{f}$ is flare number in that intervals. $T_{t}$ is total monitoring time in that interval. All the parameters are listed in Table 4. In the table, stars' names are given in the first column, observing seasons are listed in the second column. The average phases, total monitoring time in intervals of 0.10 phase length, the number of flares detected in that interval and the mean flare occurence rates in that interval are listed in the next columns,respectively.

\begin{table*}
\begin{center}
\caption{The report of the mean flare occurence rate for all flares (together with fast and slow flares) for all program stars.\label{tbl-4}}
\begin{tabular}{lccccc}
\hline										
Observing	&	Observing	&	Average	&	Observing	&	Flare	&	Flare	\\
Star	&	Season	&	Phase	&	Duration (h)	&	Number	&	Occur. Rate ($h^{-1}$)	\\
\hline\hline											
AD Leo	&	2004-2007	&	0.05	&	9.17	&	11	&	1.200	\\
AD Leo	&	2004-2007	&	0.15	&	2.74	&	5	&	1.825	\\
AD Leo	&	2004-2007	&	0.35	&	6.73	&	15	&	2.229	\\
AD Leo	&	2004-2007	&	0.45	&	4.60	&	12	&	2.610	\\
AD Leo	&	2004-2007	&	0.55	&	15.65	&	23	&	1.470	\\
AD Leo	&	2004-2007	&	0.65	&	8.19	&	18	&	2.198	\\
AD Leo	&	2004-2007	&	0.75	&	9.92	&	19	&	1.915	\\
AD Leo	&	2004-2007	&	0.85	&	5.11	&	6	&	1.174	\\
AD Leo	&	2004-2007	&	0.95	&	2.40	&	3	&	1.250	\\
\hline											
EV Lac	&	2004	&	0.05	&	3.88	&	3	&	0.773	\\
EV Lac	&	2004	&	0.15	&	3.62	&	2	&	0.553	\\
EV Lac	&	2004	&	0.25	&	8.20	&	4	&	0.488	\\
EV Lac	&	2004	&	0.45	&	0.91	&	1	&	1.099	\\
EV Lac	&	2004	&	0.55	&	4.19	&	4	&	0.955	\\
EV Lac	&	2004	&	0.75	&	5.17	&	7	&	1.354	\\
EV Lac	&	2004	&	0.85	&	3.53	&	3	&	0.850	\\
EV Lac	&	2004	&	0.95	&	8.12	&	7	&	0.862	\\
\hline											
EV Lac	&	2005	&	0.25	&	1.58	&	2	&	1.266	\\
EV Lac	&	2005	&	0.45	&	1.88	&	4	&	2.128	\\
EV Lac	&	2005	&	0.55	&	4.73	&	8	&	1.690	\\
EV Lac	&	2005	&	0.75	&	4.36	&	7	&	1.606	\\
EV Lac	&	2005	&	0.85	&	4.63	&	5	&	1.080	\\
EV Lac	&	2005	&	0.95	&	4.74	&	5	&	1.055	\\
\hline											
EV Lac	&	2006	&	0.05	&	3.51	&	3	&	0.855	\\
EV Lac	&	2006	&	0.15	&	2.35	&	2	&	0.850	\\
EV Lac	&	2006	&	0.25	&	5.01	&	8	&	1.597	\\
EV Lac	&	2006	&	0.35	&	3.65	&	7	&	1.918	\\
EV Lac	&	2006	&	0.45	&	2.01	&	3	&	1.493	\\
EV Lac	&	2006	&	0.55	&	2.17	&	3	&	1.382	\\
EV Lac	&	2006	&	0.65	&	2.10	&	3	&	1.429	\\
EV Lac	&	2006	&	0.85	&	1.33	&	1	&	0.752	\\
EV Lac	&	2006	&	0.95	&	9.54	&	5	&	0.524	\\
\hline											
V1005 Ori	&	2005-2006	&	0.05	&	3.78	&	7	&	1.850	\\
V1005 Ori	&	2005-2006	&	0.25	&	2.48	&	4	&	1.613	\\
V1005 Ori	&	2005-2006	&	0.35	&	7.79	&	6	&	0.770	\\
V1005 Ori	&	2005-2006	&	0.65	&	1.42	&	3	&	2.113	\\
V1005 Ori	&	2005-2006	&	0.85	&	4.45	&	11	&	2.472	\\
\hline											
EQ Peg	&	2004-2005	&	0.05	&	16.58	&	8	&	0.483	\\
EQ Peg	&	2004-2005	&	0.15	&	6.87	&	3	&	0.437	\\
EQ Peg	&	2004-2005	&	0.35	&	14.69	&	3	&	0.204	\\
EQ Peg	&	2004-2005	&	0.45	&	17.87	&	9	&	0.504	\\
EQ Peg	&	2004-2005	&	0.55	&	22.31	&	10	&	0.448	\\
EQ Peg	&	2004-2005	&	0.65	&	14.01	&	3	&	0.214	\\
EQ Peg	&	2004-2005	&	0.75	&	32.90	&	17	&	0.517	\\
EQ Peg	&	2004-2005	&	0.85	&	30.60	&	7	&	0.229	\\
EQ Peg	&	2004-2005	&	0.95	&	19.59	&	13	&	0.664	\\
\hline
\end{tabular}
\end{center}
\end{table*}											

The flare data obtained from EV Lac, EQ Peg, V1005 Ori and AD Leo were used for this analysis since the photometric periods are known for these stars. Total monitoring time of AD Leo is 79.11 $h$, while it is 109.63 $h$ for EV Lac. Total monitoring time of EQ Peg is 100.26$h$. In the case of V1005 Ori, it is 44.75 $h$. V1005 Ori and EQ Peg were observed in two seasons, while others were observed in three seasons. When the phase intervals of flare monitoring are examined, it is seen that flare monitoring was acquired almost every phase intervals. On the other hand, no flare was detected in some monitoring intervals. This is not unexpected case. This is because of the nature of UV Ceti type stars and flare processes. Because the flare activity is not a periodic or cyclic variation and it is hard to predict when a flare occur. This is why we did not detect any flare in some observations.

\begin{figure*}
\hspace{55mm}
\FigureFile(130mm,60mm){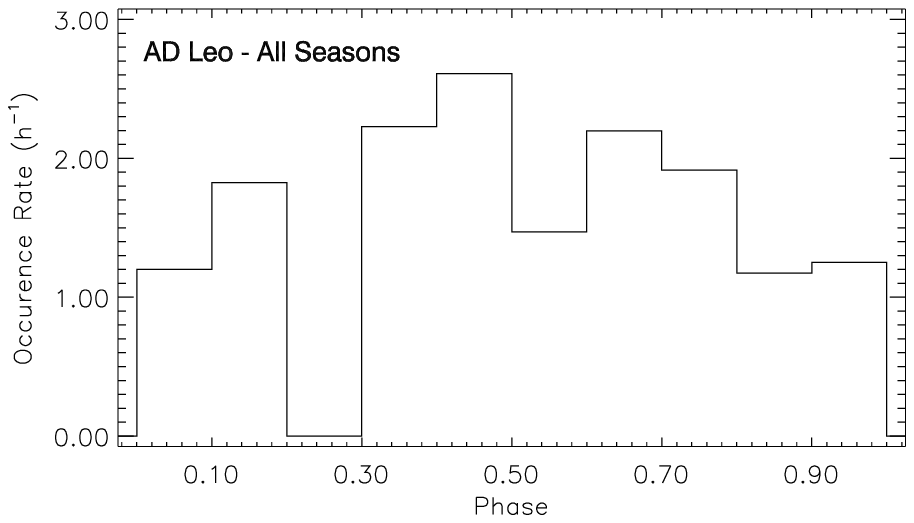}
\caption{The mean flare occurence rates versus rotational phase are demonstrated for all AD Leo flares observed in three seasons. In the figures, the lines show the histograms of mean flare occurence rates computed in intervals of 0.10 phase length.\label{fig8}}
\end{figure*}

To carry out the investigation, the seasons, in which large number of flares is detected, were chosen for each star and these seasons were only used in analyses. Thus, the distributions of detected flares can be covered almost each phase interval for each star. Nevertheless, it was seen that there was no flare some phases interval. Using the Least-Squares Method, all the histograms of $N$ were analysed with the SPSS V17.0 \citep{Gre99} software to determine the phase in which Maximum Flare Occurence Rates (hereafter MFOR) are seen. As it is seen in the figures, the histograms are not sometimes shown from $0.0$ to $1.0$ in phase. This is because the histograms and models are shown in the best view.

\begin{figure*}
\hspace{55mm}
\FigureFile(130mm,60mm){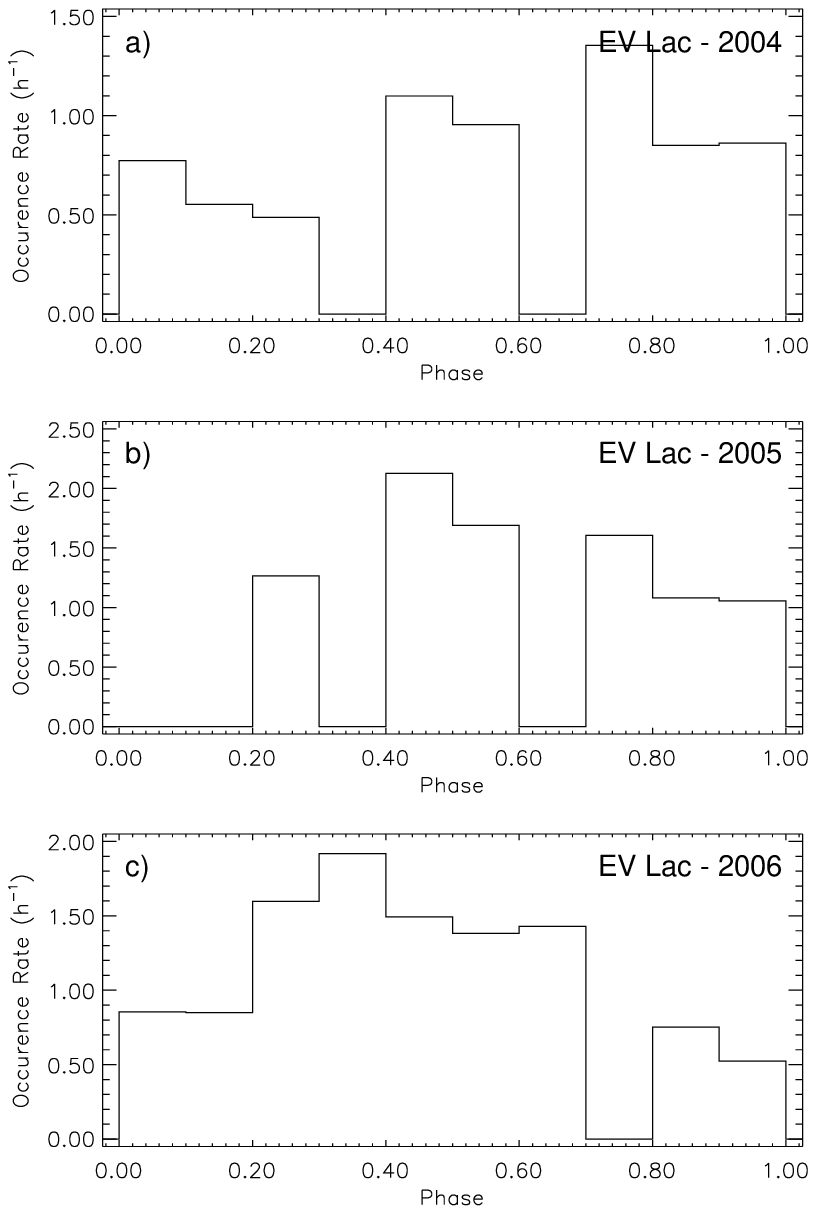}
\caption{The mean flare occurence rates versus rotational phase are demonstrated for EV Lac flares observed in three seasons. In the figures, the lines show the histograms of mean flare occurence rates computed in intervals of 0.10 phase length. The histograms are for the season 2004 (a) 2005 (b) and the season 2006 (c).\label{fig9}}
\end{figure*}

AD Leo was observed for three seasons. 119 U band flares were detected in 79.61 $h$ monitoring time. The analyses do not reveal any variability out-of-flare. Because of this, we combine all 119 flares detected in three seasons and we derived one histogram for all of them. The histogram of the mean flare occurence rates versus rotational phase for AD Leo flares are shown in Figure 8. In the figures, the mean flare occurence rates versus rotational phase are demonstrated by histogram. As it is seen from the analyses of the histogram in Figure 8, the phase of MFOR is about $0^{P}.45$ for detected flares.

\begin{figure*}
\hspace{55mm}
\FigureFile(130mm,60mm){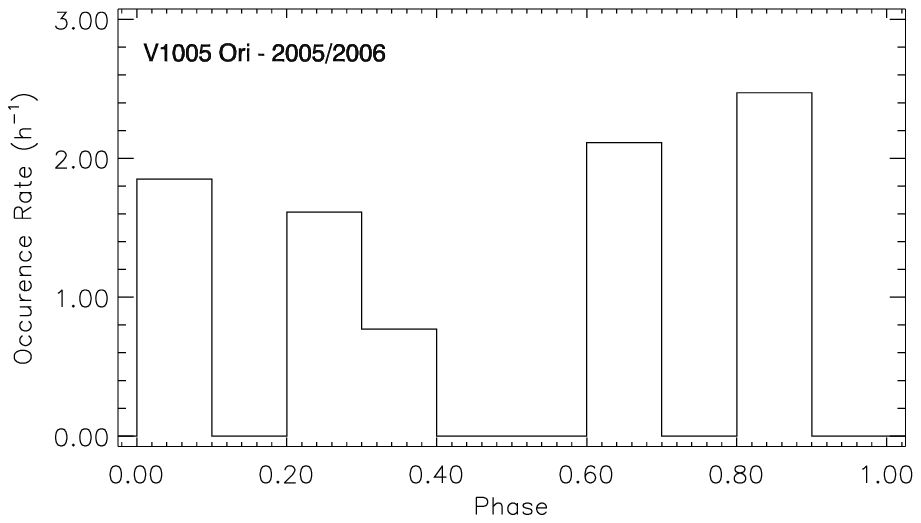}
\caption{The mean flare occurence rates versus rotational phase are demonstrated for V1005 Ori flares observed in the season 2005/2006. In the figures, the lines show the histograms of mean flare occurence rates computed in intervals of 0.10 phase length.\label{fig10}}
\end{figure*}

EV Lac was also observed for three seasons. Although monitoring time is 109.63 $h$, but 93 U band flares were detected. On the other hand, the distributions of the flares are covered almost all phases for each observing season. Because of this, we derived histograms of the distributions of the flares versus phases for each season. The histograms are shown for EV Lac flares in Figure 9. EV Lac was observed in the seasons 2004, 2005 and 2006. According to the light curves of the rotational modulation for EV Lac, the two data sets were obtained for the season 2004. The flares were not separated into two sets for analyses. This is because the minimum phases of rotational modulations for Set 1 and 2 are the same. The histograms also exhibit the same distributions, when the flares observed in 2004 are separated into two sets. Thus, one histogram is shown in Figure 9 for the season 2004. As it is seen from the analyses of the histogram in Figure 9, the phases of MFOR are $0^{P}.75$ for 2004, $0^{P}.45$ for 2005 and $0^{P}.35$ for 2006.

\begin{figure*}
\hspace{53mm}
\FigureFile(140mm,60mm){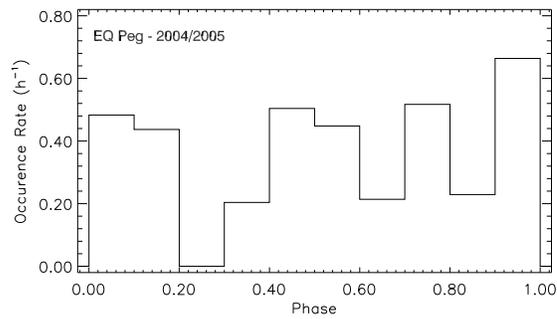}
\caption{The mean flare occurence rates versus rotational phase are demonstrated for EQ Peg flares observed in the seasons 2004 and 2005. In the figure, the lines show the histograms of mean flare occurence rates computed in intervals of 0.10 phase length.\label{fig11}}
\end{figure*}

Flare patrol of V1005 Ori was continued in the seasons of 2004/2005 and 2005/2006. 44 U band flares were detected in total 44.75 $h$ monitoring time. When we checked the phase distributions of detected flares for each season, it is seen that the phase distribution of flares detected in the season 2004/2005 is not covered all phases very well. Because of this, we only used the data of the season 2005/2006 in the analyses. The same histogram is shown in Figure 10 for the season 2005/2006 of V1005 Ori. As it is seen from the analyses of the histogram in Figure 10, the flare occurence rate of detected flares reaches maximum value in about $0^{P}.88$.

In the case of EQ Peg, observations were acquired for two seasons. Because of the similarity of the light curves in seasons 2004 and 2005, all flares of EQ Peg were analysed together and derived histogram is shown in Figure 11. The analyses showed that the phase of MFOR is about $0^{P}.95$

\subsection{Distribution of The Fast and Slow Flares According to Each Other}

In order to test the hypothesis discussed by \citet{Gur86}, according to the rule described by \citet{Dal10}, the observed flares were separated into two classes as slow and fast flares. For each program star, the data sets of both slow and fast flares were composed for each observation season. Then, the same analyses mentioned above were derived with these sets to compare distribution of both flare types according to each other.

All the parameters are listed in Table 5 for the fast flares and in Table 6 for the slow flares. All the columns of these tables are the same with Table 4. It is important to note that given values of "total monitoring time in intervals of 0.10 phase length" in these tables could be slightly different from each other and from the values given in Table 4. This is because this parameter was separately computed for each flare types and for all flares (together with fast and slow flares) for each program star.

\begin{table*}
\begin{center}
\caption{The report of the mean flare occurence rate for the fast flares for all program stars.\label{tbl-5}}
\begin{tabular}{lccccc}
\hline									
Observing	&	Observing	&	Average	&	Observing	&	Flare	&	Flare	\\
Star	&	Season	&	Phase	&	Duration (h)	&	Number	&	Occur. Rate ($h^{-1}$)	\\
\hline\hline											
AD Leo	&	2004-2006	&	0.05	&	9.09	&	3	&	0.330	\\
AD Leo	&	2004-2006	&	0.35	&	6.73	&	5	&	0.743	\\
AD Leo	&	2004-2006	&	0.45	&	3.02	&	2	&	0.662	\\
AD Leo	&	2004-2006	&	0.55	&	7.03	&	5	&	0.711	\\
AD Leo	&	2004-2006	&	0.65	&	2.07	&	2	&	0.964	\\
AD Leo	&	2004-2006	&	0.75	&	5.41	&	4	&	0.740	\\
AD Leo	&	2004-2006	&	0.95	&	6.40	&	3	&	0.469	\\
\hline											
EV Lac	&	2004	&	0.05	&	2.81	&	1	&	0.356	\\
EV Lac	&	2004	&	0.15	&	2.21	&	1	&	0.452	\\
EV Lac	&	2004	&	0.35	&	1.79	&	1	&	0.559	\\
EV Lac	&	2004	&	0.45	&	1.71	&	1	&	0.584	\\
EV Lac	&	2004	&	0.55	&	10.60	&	3	&	0.283	\\
EV Lac	&	2004	&	0.55	&	5.35	&	1	&	0.187	\\
\hline											
EV Lac	&	2006	&	0.15	&	4.06	&	2	&	0.493	\\
EV Lac	&	2006	&	0.25	&	5.01	&	2	&	0.399	\\
EV Lac	&	2006	&	0.45	&	2.33	&	1	&	0.429	\\
EV Lac	&	2006	&	0.55	&	2.17	&	1	&	0.461	\\
EV Lac	&	2006	&	0.65	&	2.06	&	1	&	0.486	\\
EV Lac	&	2006	&	0.95	&	1.59	&	1	&	0.629	\\
\hline											
V1005 Ori	&	2005-2006	&	0.05	&	5.78	&	2	&	0.346	\\
V1005 Ori	&	2005-2006	&	0.25	&	2.48	&	2	&	0.806	\\
V1005 Ori	&	2005-2006	&	0.35	&	2.96	&	2	&	0.675	\\
V1005 Ori	&	2005-2006	&	0.65	&	1.42	&	2	&	1.408	\\
V1005 Ori	&	2005-2006	&	0.85	&	4.45	&	6	&	1.348	\\
\hline											
EQ Peg	&	2004-2005	&	0.05	&	11.52	&	5	&	0.434	\\
EQ Peg	&	2004-2005	&	0.15	&	4.20	&	2	&	0.476	\\
EQ Peg	&	2004-2005	&	0.35	&	11.13	&	2	&	0.180	\\
EQ Peg	&	2004-2005	&	0.45	&	14.31	&	3	&	0.210	\\
EQ Peg	&	2004-2005	&	0.55	&	22.31	&	4	&	0.179	\\
EQ Peg	&	2004-2005	&	0.65	&	14.01	&	3	&	0.214	\\
EQ Peg	&	2004-2005	&	0.75	&	32.90	&	10	&	0.304	\\
EQ Peg	&	2004-2005	&	0.85	&	22.36	&	4	&	0.179	\\
EQ Peg	&	2004-2005	&	0.95	&	19.59	&	9	&	0.459	\\
\hline
\end{tabular}
\end{center}
\end{table*}

\begin{table*}
\begin{center}
\caption{The report of the mean flare occurence rate for the slow flares for all program stars.\label{tbl-6}}
\begin{tabular}{lccccc}
\hline											
Observing	&	Observing	&	Average	&	Observing	&	Flare	&	Flare	\\
Star	&	Season	&	Phase	&	Duration (h)	&	Number	&	Occur. Rate ($h^{-1}$)	\\
\hline\hline											
AD Leo	&	2004-2005	&	0.05	&	10.36	&	8	&	0.772	\\
AD Leo	&	2004-2005	&	0.15	&	2.74	&	5	&	1.825	\\
AD Leo	&	2004-2005	&	0.35	&	5.03	&	10	&	1.990	\\
AD Leo	&	2004-2005	&	0.45	&	6.31	&	17	&	2.694	\\
AD Leo	&	2004-2005	&	0.55	&	9.94	&	17	&	1.710	\\
AD Leo	&	2005-2006	&	0.65	&	6.68	&	9	&	1.347	\\
AD Leo	&	2005-2006	&	0.75	&	9.92	&	15	&	1.512	\\
AD Leo	&	2005-2006	&	0.85	&	5.11	&	3	&	0.587	\\
AD Leo	&	2005-2006	&	0.95	&	2.23	&	1	&	0.448	\\
\hline											
EV Lac	&	2004	&	0.05	&	3.88	&	2	&	0.515	\\
EV Lac	&	2004	&	0.15	&	2.21	&	1	&	0.452	\\
EV Lac	&	2004	&	0.25	&	3.03	&	2	&	0.659	\\
EV Lac	&	2004	&	0.35	&	2.69	&	2	&	0.743	\\
EV Lac	&	2004	&	0.45	&	0.91	&	1	&	1.099	\\
EV Lac	&	2004	&	0.55	&	1.08	&	1	&	0.926	\\
EV Lac	&	2004	&	0.75	&	5.17	&	3	&	0.580	\\
EV Lac	&	2004	&	0.85	&	3.53	&	2	&	0.567	\\
EV Lac	&	2004	&	0.95	&	8.13	&	3	&	0.369	\\
\hline											
EV Lac	&	2006	&	0.05	&	3.51	&	3	&	0.855	\\
EV Lac	&	2006	&	0.25	&	5.01	&	6	&	1.198	\\
EV Lac	&	2006	&	0.35	&	4.94	&	7	&	1.417	\\
EV Lac	&	2006	&	0.45	&	1.70	&	2	&	1.176	\\
EV Lac	&	2006	&	0.55	&	2.17	&	2	&	0.922	\\
EV Lac	&	2006	&	0.65	&	2.10	&	2	&	0.952	\\
EV Lac	&	2006	&	0.85	&	1.33	&	1	&	0.752	\\
EV Lac	&	2006	&	0.95	&	9.55	&	4	&	0.419	\\
\hline											
V1005 Ori	&	2005-2006	&	0.05	&	2.45	&	3	&	1.222	\\
V1005 Ori	&	2005-2006	&	0.25	&	2.48	&	2	&	0.806	\\
V1005 Ori	&	2005-2006	&	0.35	&	7.80	&	4	&	0.513	\\
V1005 Ori	&	2005-2006	&	0.65	&	1.26	&	1	&	0.792	\\
V1005 Ori	&	2005-2006	&	0.85	&	4.45	&	5	&	1.124	\\
V1005 Ori	&	2005-2006	&	0.95	&	2.06	&	2	&	0.970	\\
\hline											
EQ Peg	&	2004-2005	&	0.05	&	12.40	&	3	&	0.242	\\
EQ Peg	&	2004-2005	&	0.15	&	2.67	&	1	&	0.375	\\
EQ Peg	&	2004-2005	&	0.35	&	3.56	&	1	&	0.281	\\
EQ Peg	&	2004-2005	&	0.45	&	17.86	&	6	&	0.336	\\
EQ Peg	&	2004-2005	&	0.55	&	15.31	&	6	&	0.392	\\
EQ Peg	&	2004-2005	&	0.75	&	25.09	&	7	&	0.279	\\
EQ Peg	&	2004-2005	&	0.85	&	10.68	&	3	&	0.281	\\
EQ Peg	&	2004-2005	&	0.95	&	16.60	&	4	&	0.241	\\
\hline
\end{tabular}
\end{center}
\end{table*}											

\begin{figure*}
\hspace{62mm}
\FigureFile(130mm,60mm){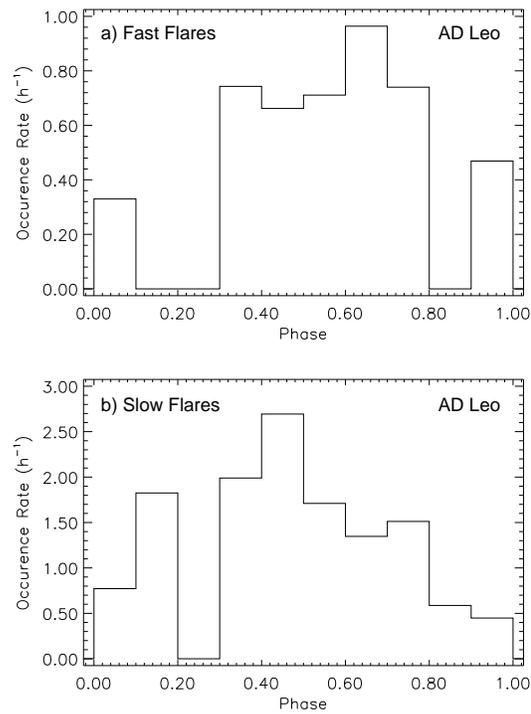}
\caption{The histograms of the mean flare occurrence rates versus rotational phase are for both slow and fast flares of AD Leo. a) Fast flares observed in three seasons. b) Slow flares observed in three seasons.\label{fig12}}
\end{figure*}

All the flares of AD Leo detected in three seasons were again combined for this analysis. The same histograms were derived for both the fast and slow flares of AD Leo. They are shown in Figure 12. As it is seen from the analyses of the histogram in the figure, the phase of MFOR is $0^{P}.58$, while it is $0^{P}.41$ for the slow flares. There is a difference of $0^{P}.17$ between two types. According to \citet{Gur86}, it is expected that there should be a difference of $0^{P}.50$ between them.

\begin{figure*}
\hspace{30mm}
\vspace{5mm}
\FigureFile(120mm,60mm){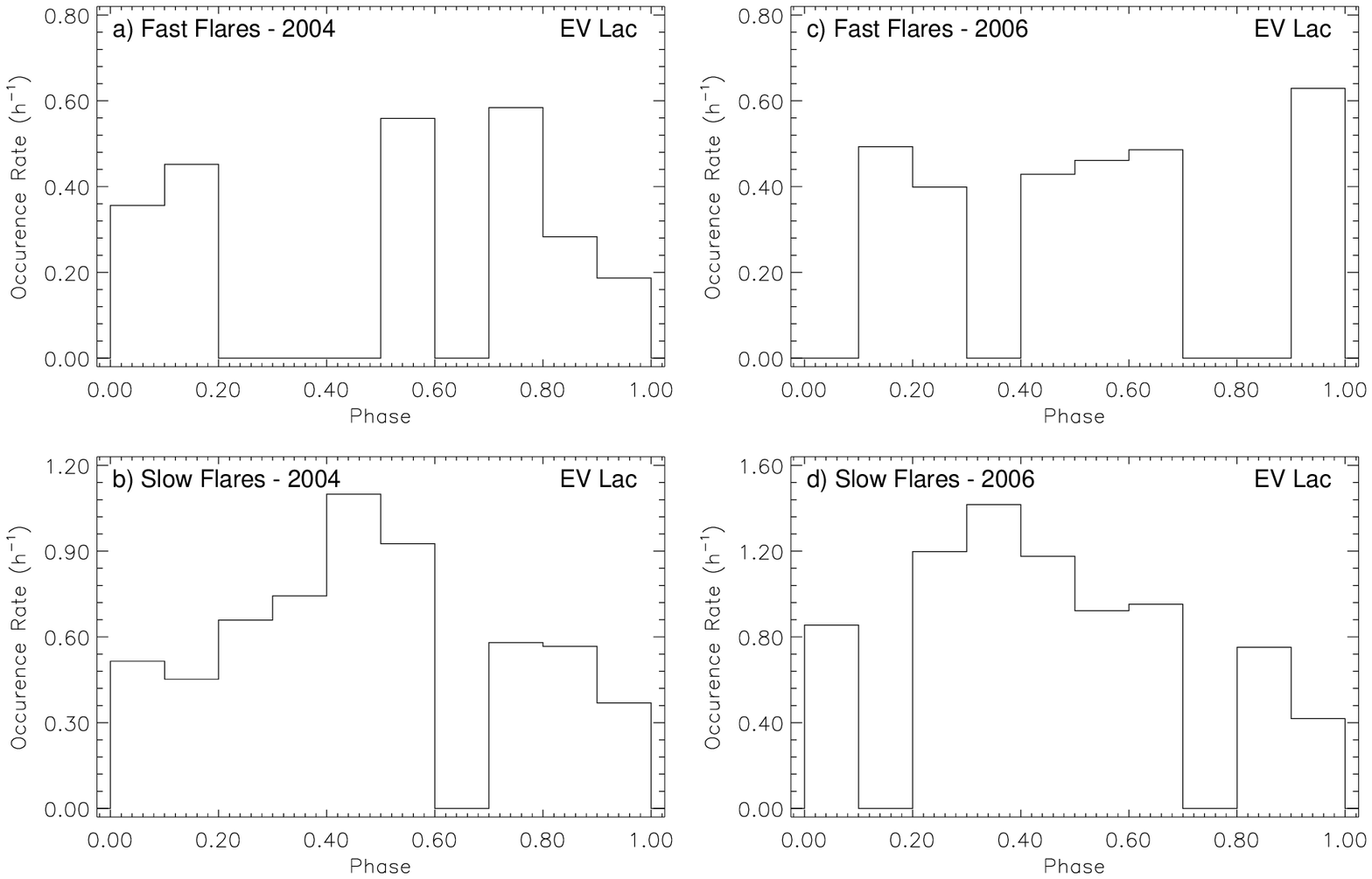}
\caption{The histograms of the mean flare occurrence rates versus rotational phase are for both slow and fast flares of EV Lac. a) Fast flares of the season 2004. b) Slow flares of the season 2004. c) Fast flares of the season 2006. d) Slow flares of the season 2006.\label{fig13}}
\end{figure*}

In the case of EV Lac, it was seen that the phase distribution of the fast flares is not enough to compare it with slow flares for the season 2005. This is must be because the frequency of the fast flares is not as high as that of the slow flares, as mentioned by \citet{Dal10}. We only compared them for the season 2004 and 2006. All histograms of EV Lac are shown in Figure 13. The phase of MFOR for the fast flares is $0^{P}.45$, while it is $0^{P}.49$ for the slow flares in the season 2004. The difference between the phases of MFOR is $0^{P}.04$ for two types in this season. The phase of MFOR is $0^{P}.87$ for the fast flares, while it is $0^{P}.36$ for the slow flares in the season 2006. The difference between the phases of MFOR is $0^{P}.51$ for both types in the season 2006, as expected.

\begin{figure*}
\hspace{60mm}
\vspace{5mm}
\FigureFile(130mm,60mm){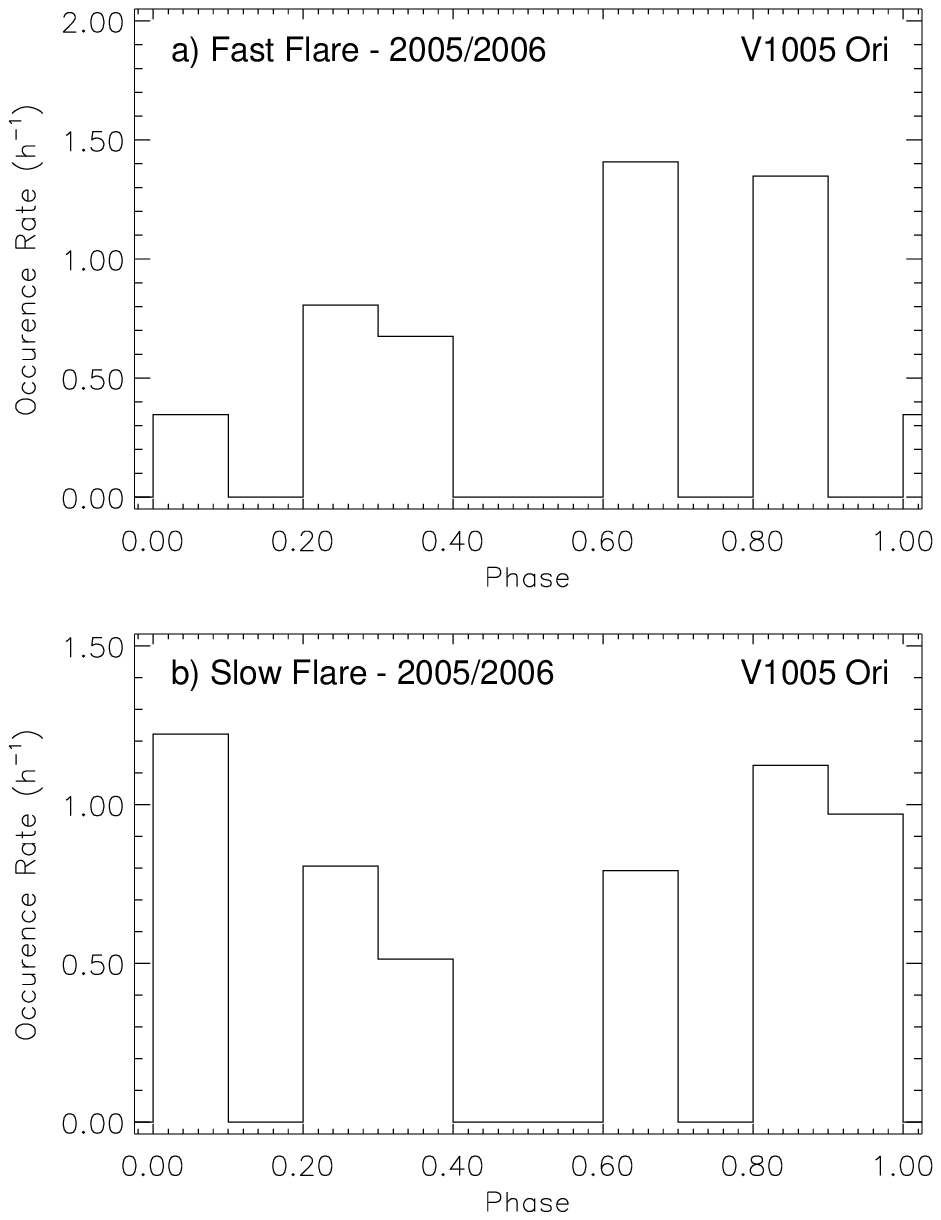}
\caption{The histograms of the mean flare occurrence rates versus rotational phase are for both slow and fast flares of V1005 Ori. a) Fast flares of the season 2005/2006. b) Slow flares of the season 2005/2006.\label{fig14}}
\end{figure*}

In the case of V1005 Ori, comparison could be done for the season 2005/2006. The histograms of V1005 Ori are shown in Figure 14. As it is seen from the analyses of the histogram in the figure, the phase of MFOR for the fast flares is $0^{P}.64$, while it is $0^{P}.95$ for the slow flares. There is a difference of $0^{P}.31$ between two types.

\begin{figure*}
\hspace{60mm}
\vspace{5mm}
\FigureFile(130mm,60mm){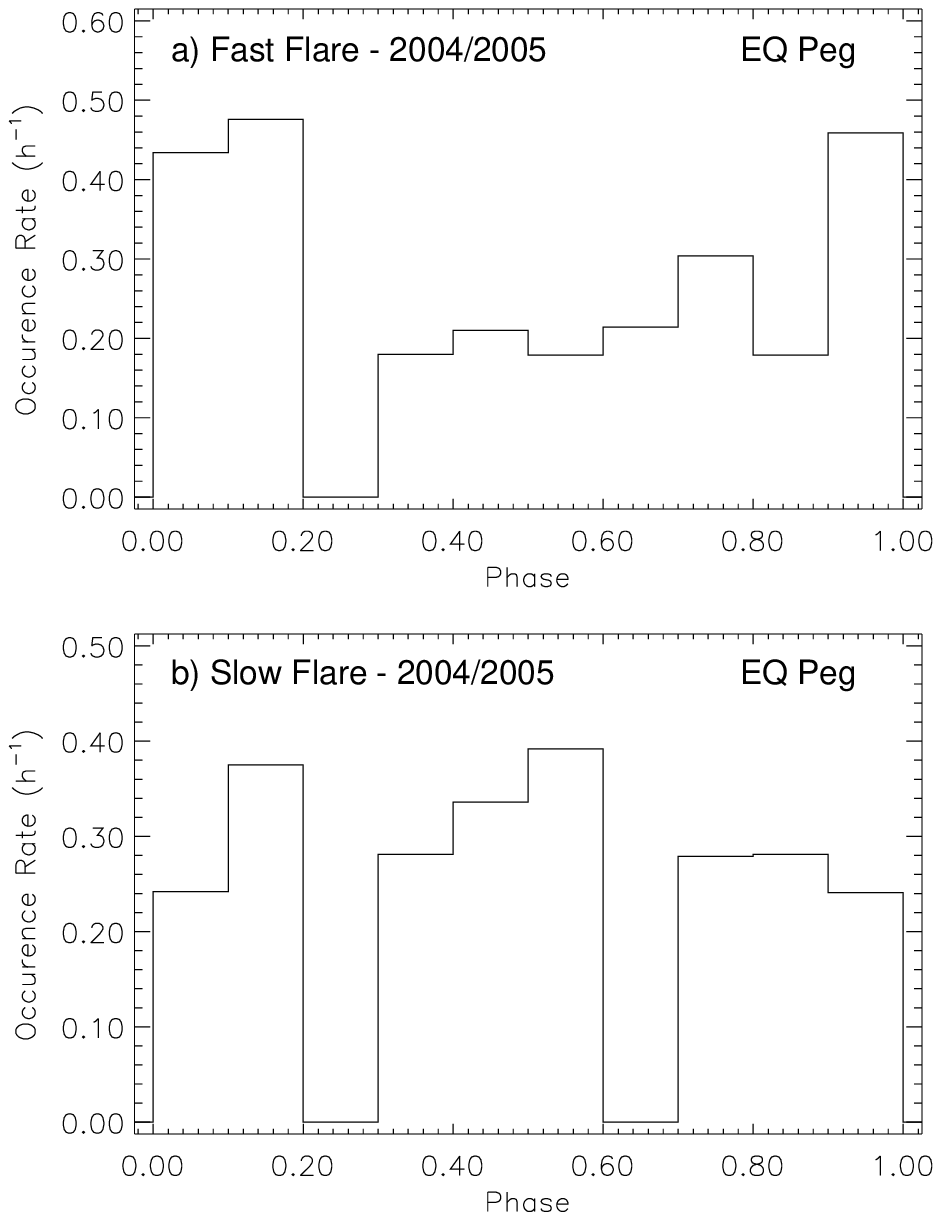}
\caption{The histograms of the mean flare occurrence rates versus rotational phase are for both slow and fast flares of EQ Peg. a) Fast flares of the seasons 2004 and 2005. b) Slow flares of the seasons 2004 and 2005.\label{fig15}}
\end{figure*}

The same comparison was done for the seasons 2004 and 2005 for EQ Peq. The histograms of EQ Peg are shown in Figure 15. As it is seen from the analyses of the histogram in the figure, the phase of MFOR for the fast flares is about $0^{P}.15$, while it is $0^{P}.55$ for the slow flares. There is a difference of $0^{P}.40$ between two types. This value is an acceptable value and close to the expected value according to the hypothesis discussed by \citet{Gur86}.

\section{Results and Discussion}

\subsection{Stellar Rotational Modulation and Stellar Spot Activity}

Most of the UV Ceti type stars are full convective red dwarfs with sudden-high energy emitting. As it can be seen in the literature, BY Dra Syndrome at out-of-flares is seen in a few stars among 463 flare stars catalogued by \citet{Ger99}. EV Lac and V1005 Ori can be given as two examples because the studies in the literature and this study indicate that both stars show the variation due to rotational modulation at out-of-flares.

In the case of EV Lac, the time series analyses show that the period of rotational modulation found for each data set is range from $4^{d}.330$ to $4^{d}.378$. The periods found are similar to those found by \citet{Pet83} and \citet{Mah81}. Although the periods found for each season are a little bit different, this difference is relatively small. When the amplitudes of the light curves are examined for EV Lac, the amplitude of this variation was dramatically decreasing from the year 2004 to 2005, while the amplitude was clearly larger than ever in this study. However, the mean average of brightness in the light curves was slowly decreasing from the year 2004 to 2006. The minima phases of the light curves for the three seasons were computed and, it was found as $0^{P}.62$ for the season 2004, $0^{P}.54$ for 2005 and $0^{P}.60$ for 2006.

In the case of V1005 Ori, the periods of the rotational modulation for each season are range from $4^{d}.419$ to $4^{d}.429$. In the literature, \citet{Bop78} found four possible periods varied from $1^{d}.883$ to $2^{d}.199$. On the other hand, \citet{Byr84} found a period of $4^{d}.565$. As it is seen, the periods found in this study are close to the period found by \citet{Byr84}. When the amplitudes of the light curves were examined, the amplitude observed in the season of 2004/2005 was so smaller than the ones observed in the previous and later seasons that there was no minimum in the light curve.  Although the mean average of brightness in the light curves was not changing, the minimum phases of the light curves were varying. The minimum phase of the curve for the season 2004/2005 was about $0^{P}.34$ and about $0^{P}.66$ for the season 2006/2007. It is hard to say that the minimum phase of the light curves for the season 2005/2006 was about $0^{P}.78$.

The case of AD Leo is different from the other two stars. The time series analyses do not show any regular variation over the $3\sigma$ level in one season. On the other hand, the mean brightness levels were increased a value of $0^{m}.01$ from the first season to the second and a value of $0^{m}.02$ from the second to the last season. This can be because of the stellar polar spots. If the literature is considered, the stellar spots can be carried to polar regions in the case of rapid rotation in the young stars \citep{Sch92, Sch96}. According to \citet{Mon01}, AD Leo is at the age of 200 Myr. The range of equatorial rotational velocity ($vsini$) given in the literature is between $5$ - $5.8$ and $9.0$ $kms^{-1}$ for AD Leo \citep{Mar92, Pet91}. Besides, considering these values of $vsini$, the real rotational velocities must be larger than these values. If both the age and equatorial rotational velocity value parameters in these papers are considered, according to \citet{Sch92} and \citet{Sch96}, some spots might be located on the polars for AD Leo. In fact, \citet{Pet92} indicate that BY Dra had spotted area near polar region, which was stable for 14 years and EV Lac has a similar area for 10 years. If the studies made by \citet{Spi86} and \citet{Pan93} are considered, AD Leo might sometimes show rotational modulation due to the spotted area occurring near the equatorial regions. On the other hand, there is another probability. If the colour index of V-R is considered, it is seen that the star gets bluer from a season to next one, when the star get brighter. Besides, no amplitude is seen in the light curves. These can be some indicators that all the surface of the star is covered by cool spots and the efficiency of the spots gets weaker from one season to next one.

The colour curves of both EV Lac and V1005 Ori sometimes exhibit a clear colour excess around the minimum phases of the light curves for some observing seasons. This can be an indicator of some bright areas such as faculae on the surface of these young stars. The effects of the bright areas such as faculae can be seen in the variations of B-V and sometimes V-R colour, while these effects are not seen in the variations in the light curves of BVR bands due to cool spots. The cool spots are more efficient in the light curves of B, especially V and R bands. The same effect is seen in the variations caused by the flare activity. Although there is some small effects or no effect of the flare activity in V and R bands, but there is some clear variations in U band light and U-B colour curves. In this study, we observed the stars in U band to investigate the variations out-of-flares. In a sense, U band observations is used to control whether there is any flare activity in the observing durations. If there is some variations in U band light and U-B colour, we did not used the observation to investigate the variation out-of-flares.

\citet{Nor07} showed that EQ Peg has a variability with the period of $1^{d}.0664$. In this study, the time series analyses supported this period. According to our analyses, EQ Peg exhibits short-term variability with the period of $1^{d}.0608$. However, it is seen that there is not any variability in the colour indexes. Analysing the light curve of EQ Peg, it was found as $0^{P}.32$ for the minimum phase of the rotational modulation.

\subsection{The Relation Between Stellar Spot and Flare Activities}

There are many studies about whether the flares of UV Ceti type stars showing BY Dra Syndrome are occurring at the same longitudes of stellar spots, or not. Having the same longitudes of flare and spots is an expected case for these stars, because solar flares are mostly occurring in the active regions, where spots are located on the Sun \citep{Ben10}. In the respect of Stellar-Solar Connection, a result of the $Ca$ $II$ $H\&K$ Project of Mount Wilson Observatory \citep{Wil78, Bal95}, if the areas of flares and spots are related on the Sun, the same case might be expected for the stars. In fact, \citet{Mon96} have found some evidence to demonstrate this relations. Besides, \citet{Let97} have found a variations of both the rotational modulation and the phase distribution of flare occurence rates in the same way for the observations in the year 1970. On the other hand, no clear relation between stellar flares and spots has been found by \citet{Bop74, Pet83}. However, \citet{Pet83} did not draw firm conclusions because of being a non-uniqueness problem.

In this study, the flare occurence rates, the ratio of flare number to monitoring time, were computed in intervals of 0.10 phase length as the same method used by \citet{Let97} with just one difference. The flare maximum times were used to compute the phases due to main energy emitting in this part of the flare light curves. We observed AD Leo for 79.61 $h$ and detected 119 flares in three seasons. EV Lac was observed 109.63 $h$ and 93 flares were detected in three seasons. V1005 Ori was observed for 44.75 $h$ and 44 flares were detected in two seasons. EQ Peg was observed for 100.26 $h$ and 73 U band flare were detected. Since no rotational modulation was found to compare for AD Leo, all the flares detected in three season were combined in order to just find whether there is any phase, in which the flare occurence rate gets a peak. On the other hand, we examined flare phase distributions for each season for both EV Lac and V1005 Ori. In the case of these stars, if the distribution of flares did not cover almost all phases in an observing season of a star, the season is neglected for the comparison of flare and spot activity. Consequently, for both EV Lac and V1005 Ori, we chose the seasons, in which the best flare distributions were obtained. Thus, we only used the seasons, in which there is enough data to get reliable conclusions about flare occurrence distributions. In addition, to determine the phases of MFOR, all the distributions were modelled with the polynomial function. Resolving these models, maximum flare occurrence rates and their phase were found for all program stars.

In the case of EV Lac, no relation is seen between the minimum phase of the rotational modulation and the phase, in which flare activity reaches the MFOR. The minimum phase of the rotational modulation observed in the season 2004 is $0^{P}.62$, while the phase of MFOR is $0^{P}.75$. The minimum phase of rotational modulation is $0^{P}.54$, while the flare occurrence rate reaches maximum level in about the phase of $0^{P}.45$ for the season 2005. In the last season of EV Lac, rotational modulation minimum is seen in $0^{P}.60$, as MFOR is in $0^{P}.35$. In the case of V1005 Ori, there is enough data in only one season to compare. As it is seen, the minimum phase of rotational modulation is $0^{P}.78$, while phase of MFOR is about $0^{P}.87$ for the season 2005/2006. In this study, the time series analyses indicated that AD Leo does not have any rotational modulation. Therefore, any minimum time could not have been determined from the observations of three seasons for AD Leo. Because of this, we could not compare the rotational modulation with flare activity in the case of AD Leo. On the other hand, using combined data of three seasons, we found that the MFOR is seen in $0^{P}.45$. This phases was computed with using the ephemeris given in Equation (1) taken from \citet{Pan93}. The time series analyses do not show any short-term variation in the light curves of AD Leo. Because of this, we waited that there is no any phase, in which the flare activity gets higher levels. On the other hand, as it is seen from the histogram and its Normal Gaussian model for AD Leo, there is a phase for MFOR. Considering the phase of MFOR, the active region(s) in some particular part of the surface can be more active than the others on the surface of the star. Considering the light and colour curves of AD Leo, almost all surface of the star may be covered by stellar spots, while it is seen that some region(s) in the surface of the star can be more active than the remainder of the surface. In the case of EQ Peg, the minimum phase of the rotational modulation is $0^{P}.32$, while the phase of MFOR is $0^{P}.95$.

The results acquired from EV Lac and V1005 Ori demonstrated that flare activity can reach high levels at almost the same longitudes, in which stellar spots occur. On the other hand, there is a considerable difference between the phases of stellar spot and MFOR for the observing season 2007 of EV Lac. In conclusion, it is seen that there is a longitudinal relation between stellar spot and flare activities in general manner. Nevertheless, there are some differences and this makes difficult to do a definite conclusion. Moreover, in the case of EQ Peg, the MFOR gets the minimum towards the minimum phase of the rotational modulation. All these cases can be because of a dynamo which is working in the red dwarf stars. In spite of the Sun, red dwarf stars are mostly known to have a different dynamo because of full convective outer atmosphere. However, in the last years, some studies showed that flares on the Sun do not have to be located upon the spotted areas on the Sun \citep{Bor07}. In addition, it should be kept in mind that most of the studies have been done with using the data obtained from white-light flare observations, but a white-light flare does not have to occur in a flare process. Recent studies have shown that non white-light flares may be so common in UV Ceti-type stars as they are in the Sun \citep{Cre04, Cre06}. In this point, it can be mentioned that the analyses of data obtained from only white-light flare observations are not sufficiently qualified. For instance, \citet{Gar03} found some flares occurring in the same active area with other activity patterns with using simultaneous observations.

\subsection{Phase Distribution of The Fast And Slow Flares}

Using the inverse Compton event, \citet{Gur86} developed a hypothesis called Fast Electron Hypothesis, in which red dwarfs generate only fast flares on their surface. On the other hand, according to the flare region on the surface of the star in respect to direction of observer, the shapes of the flare light variations can be seen like a slow flare \citep{Gur86}. If the scenario in this hypothesis is working, it is expected that the fast and slow flares should collected into two phases in the light curves of UV Ceti type stars showing BY Dra Syndrome. It is also expected that these two phases are separated from each other with intervals of $0^{P}.50$ in phase.

In this study, according to the rule described by \cite{Dal10}, the flares are classified as fast and slow flares. Then the phase distributions of fast flares were compared with the phases of slow flares in order to find out whether there is any separation as expected in this respect. When the phases of both fast and slow flares are examined one by one, it is clear that both of them can occur in any phase. To reach a definite result, the phase distributions of both fast and slow flares are statistically investigated. 

As it is stated in the previous section, if the distribution of flares did not cover almost all phases in an observing season of a star, the season is neglected for that star. Consequently, we chose the seasons, in which there is enough data to get reliable conclusions about flare occurrence distributions for both fast and slow flares. In the case of AD Leo and EQ Peg, we combined all the fast flares of three seasons as we made for the slow flares. For both fast and slow flares, using Equation (5), the number of flares occurring per an hour in intervals of 0.10 phase length was computed. The obtained occurrence rates for both fast and slow flares are shown by histograms in Figures 12, 13, 14 and 15. Once again, all the distributions were modelled with the polynomial function. Resolving these models, maximum flare occurrence rates and their phase of both slow and fast flares were found for all program stars.

In the case of AD Leo, the analyses show that both fast and slow flares have a difference of $0^{P}.17$ between the phases, in which flare occurrence rates in intervals of 0.10 phase length reach maximum amplitudes. The same difference is $0^{P}.05$ for EV Lac in the season of 2004. Although these differences are acceptable as low values according to Fast Electron Hypothesis, the difference seen in the season of 2006 is $0^{P}.50$ for EV Lac. This value is the expected value in respect of Fast Electron Hypothesis. In the case of V1005 Ori, slow and fast flares could be compared only for the season of 2005/2006. The result is that both fast and slow flares have a difference of $0^{P}.30$ between the phases of maximum flare occurrence rates. In the case of EQ Peg, the phase difference between MFORs of slow and fast flares is about $0^{P}.40$. The value obtained from EQ Peg is also the expected value in respect of Fast Electron Hypothesis. It should be noted that in the case of EQ Peg, it is seen just one clear peak for the distribution of MFOR for the fast flares, while there are several peaks for the slow flares.

As it is seen from the analyses, both the fast and the slow flares sometimes the same longitudinal distributions and sometimes different. This makes difficult to say that there is a regular longitudinal division between these two types of flares as expected according to \citet{Gur86}. This means that, when a slow flare is observed, it does not have to be a fast flare occurred on the opposite side of the star in respect to observer direction.

\section*{Acknowledgments}
The authors acknowledge generous allotments of observing time at the Ege University Observatory. We thank both Dr. Hayal Boyac{\i}o\v{g}lu, who gave us important suggestions about statistical analyses, and Professor M. Can Akan, who gave us valuable suggestions that improved the language of the paper. We also thank the referee for useful comments that have contributed to the improvement of the paper. We finally thank the Ege University Research Found Council for supporting this work through grant no. 2005/FEN/051.


\begin{thebibliography}{63}
\bibitem[Amado et al. (2001)]{Ama01} Amado, P. J., Zboril, M., Butler, C. J. \& Byrne, P. B., 2001, CoSka, 31, 13
\bibitem[Anderson (1979)]{And79} Anderson, C. M., 1979, \pasp, 91, 202
\bibitem[Baliunas et al. (1995)]{Bal95} Baliunas, S.L., Donahue, R.A., Soon, W.H., Horne, J.H., Frazer, J., Woodard-Eklund, L., Bradford, M., Rao, L.M., Wilson, O.C., Zhang, Q. \& 17 coauthors, 1995, \apj, 438, 269
\bibitem[Benz \& G\"{u}del (2010)]{Ben10} Benz, A. O. \& G\"{u}del, M., 2010, ARA\&A, 48, 241
\bibitem[Bopp (1974)]{Bop74} Bopp, B. W., 1974, \apj, 193, 389
\bibitem[Bopp \& Espanak (1977)]{Bop77} Bopp, B.W. \& Espenak, F., 1977, \aj, 82, 916
\bibitem[Bopp \& Evans (1973)]{Bop73} Bopp, B. W. \& Evans, D. S., 1973, \mnras, 164, 343
\bibitem[Bopp et al. (1978)]{Bop78} Bopp, B. W., Torres, C. A. O., Busko, I. C. \& Quast, G. R., 1978, IBVS, No.1443, 1
\bibitem[Borovik et al. (2007)]{Bor07} Borovik, A. V., Myachin \& D. Yu., 2007, ASPC, 368, 411
\bibitem[Byrne et al. (1984)]{Byr84} Byrne, P. B., Doyle, J. G. \& Butler, C. J., 1984, \mnras, 206, 907
\bibitem[Chugainov (1974)]{Chu74} Chugainov, P. F., 1974, IzKry, 52, 3
\bibitem[Crespo-Chac\'{o}n et al. (2004)]{Cre04} Crespo-Chac\'{o}n, In\'{e}s, Montes, D., Fern\'{a}ndez-Figueroa, M. J., L\'{o}pez-Santiago, J., Garc\'{i}a-Alvarez, D., Foing, B. H., 2004, Ap\&SS, 292, 697
\bibitem[Crespo-Chac\'{o}n et al. (2006)]{Cre06} Crespo-Chac\'{o}n, I., Montes, D., Garc\'{i}a-Alvarez, D., Fern\'{a}ndez-Figueroa, M. J., L\'{o}pez-Santiago, J., Foing, B. H., 2006, \aap, 452, 987
\bibitem[Dal \& Evren (2010)]{Dal10} Dal, H.A. \& Evren, S., 2010, \aj, 140, 483
\bibitem[Dal \& Evren (2011)]{Dal11} Dal, H.A. \& Evren, S., 2011, \aj, 141, 33
\bibitem[Eggen (1965)]{Egg65} Eggen, O. J., 1965, Obs, 85, 191
\bibitem[Fleming et al. (1995)]{Fle95} Fleming, Th. A., Schmitt, J. H. M. M., Giampapa, M. S., 1995, \apj, 450, 401
\bibitem[Fossi et al. (1995)]{Fos95} Fossi, B. C. M., Landini, M., Fruscione, A., \& Dupuis, J., 1995, \apj, 449, 376
\bibitem[Foster (1995)]{Fos95} Foster, G., 1995, \aj, 109, 1889
\bibitem[Friedemann \& Guertler (1975)]{Fri75} Friedemann, C. \& Guertler, J., 1975, AN, 296, 125
\bibitem[Garc\'{i}a-Alvarez et al. (2003)]{Gar03}  Garc\'{i}a-Alvarez, D., Foing, B. H., Montes, D., Oliveira, J., Doyle, J. G., Messina, S., Lanza, A. F., Rodon \'{o}, M., Abbott, J., Ash, T. D. C., Baldry, I. K., Bedding, T. R., Buckley, D. A. H., Cami, J., Cao, H., Catala, C., Cheng, K. P., Domiciano de Souza, A., Jr., Donati, J.-F., Hubert, A.-M., Janot-Pacheco, E., Hao, J. X., Kaper, L., Kaufer, A., Leister, N. V., Neff, J. E., Neiner, C., Orlando, S., O'Toole, S. J., Sch \"{a} fer, D., Smartt, S. J., Stahl, O., Telting, J., Tubbesing, S., 2003, \aap, 397, 285
\bibitem[Gershberg et al. (1999)]{Ger99} Gershberg, R. E., Katsova, M. M., Lovkaya, M. N., Terebizh, A. V. \& Shakhovskaya, N. I., 1999, \aaps, 139, 555
\bibitem[Green et al. (1999)]{Gre99} Green, S. B., Salkind, N. J., Akey, T. M., 1999, "Using SPSS for Windows: Analyzing and Understanding Data", Upper Saddle River, N.J., London Prentice Hall Press, P.50
\bibitem[Gurzadian (1965)]{Gur65} Gurzadian, G. A., 1965, Ap, 1, 170
\bibitem[Gurzadian (1986)]{Gur86} Gurzadian, G. A., 1986, \apss, 125, 127
\bibitem[Gurzadian (1988)]{Gur88} Gurzadian, G. A. 1988, ApJ, 332, 183
\bibitem[Haisch et al. (1987)]{Hai87} Haisch, B. M., Butler, C. J., Doyle, J. G., \& Rodon\'{o}, M., 1987, A\&A, 181, 96
\bibitem[Hardie (1962)]{Har62} Hardie R.H., 1962, in Astronomical Techniques, ed.W.A.Hiltner (Chicago: Univ. Chicago Press), 178
\bibitem[Joy (1947)]{Joy47} Joy, A. H., 1947, \apj, 105, 96
\bibitem[Joy \& Abt (1974)]{Joy74} Joy, A. H. \& Abt, H. A., 1974, \apjs, 28, 1
\bibitem[Kleinman et al. (1987)]{Kle87} Kleinman, S. J., Sandmann, W. H. \& Ambruster, C. W., 1987, IBVS, No.3031, 1
\bibitem[Kron (1952)]{Kro52} Kron, G. E., 1952, \apj, 115, 301
\bibitem[Kukarin (1969)]{Kuk69} Kukarin, B. V., 1969, "in General Catologue of Variable Stars", 3d ed., Moscow Sternberg Astronomical Institute
\bibitem[Kunkel (1975)]{Kun75} Kunkel, W. E., 1975, IAUS, 67, 15
\bibitem[Landolt (1983)]{Lan83} Landolt, A. U., 1983, \aj, 88, 439
\bibitem[Landolt (1992)]{Lan92} Landolt, A. U., 1992, \aj, 104, 340
\bibitem[Leto et al. (1997)]{Let97} Leto, G., Pagano, I., Buemi, C. S. \& Rodon\'{o}, M., 1997, \aap, 327, 1114
\bibitem[Mahmoud \& Ol\'{a}h (1981)]{Mah81} Mahmoud, F. M. \& Ol\'{a}h, K., 1981, IBVS, No.1943, 1
\bibitem[Marcy \& Chen (1992)]{Mar92} Marcy, G. W. \& Chen, G. H., 1992, \apj, 390, 550
\bibitem[Mazeh et al. (2001)]{Maz01} Mazeh, T., Latham, D. W., Goldberg, E., Torres, G., Stefanik, R. P., Henry, T.J., Zucker, S., Gnat, O., Ofek, E. O., 2001, MNRAS, 325, 343
\bibitem[Montes et al. (1996)]{Mon96} Montes, D., Sanz-Forcada, J., Fernandez-Figueroa, M. J., Lorente, R., 1996, \aap, 310, 29
\bibitem[Montes et al. (2001)]{Mon01} Montes, D., L\'{o}pez-Santiago, J., G\'{a}lvez, M. C., Fern\'{a}ndez-Figueroa, M. J., De Castro, E. \& Cornide, M., 2001, \mnras, 328, 45
\bibitem[Mullan (1974)]{Mul74} Mullan, D. J., 1974, \apj, 192, 149
\bibitem[Norton et al. (2007)]{Nor07} Norton, A. J., Wheatley, P. J., West, R. G., Haswell, C. A., Street, R. A., Collier Cameron, A., Christian, D. J., Clarkson, W. I., Enoch, B., Gallaway, M., Hellier, C., Horne, K., Irwin, J., Kane, S. R., Lister, T. A., Nicholas, J. P., Parley, N., Pollacco, D., Ryans, R., Skillen, I., Wilson, D. M., 2007, A\&A, 467, 785
\bibitem[Panov (1993)]{Pan93} Panov, K. P., 1993, IBVS, No.3936, 1
\bibitem[Pettersen (1980)]{Pet80} Pettersen, B. R., 1980, \aj, 85, 871
\bibitem[Pettersen (1991)]{Pet91} Pettersen, B. R., 1991, MmSAI, 62, 217
\bibitem[Pettersen et al. (1983)]{Pet83} Pettersen, B. R., Kern, G. A. \& Evans, D. S., 1983, \aap, 123, 184
\bibitem[Pettersen et al. (1984)]{Pet84} Pettersen, B. R., Coleman, L. A., Evans, D. S., 1984, \apj, 282, 214
\bibitem[Pettersen et al. (1992)]{Pet92} Pettersen, B. R., Ol\'{a}h, K. \& Sandmann, W. H., 1992, \aaps, 96, 497
\bibitem[Robrade et al. (2004)]{Rob04} Robrade, J., Ness, J. -U. and Schmitt, J. H. M. M., 2004, \aap, 413, 317
\bibitem[Rodon\'{o} (1978)]{Rod78} Rodon\'{o}, M., 1978, \aap, 66, 175
\bibitem[Scargle (1982)]{Sca82} Scargle, J. D., 1982, \apj, 263, 835
\bibitem[Sch\"{u}ssler \& Solanki (1992)]{Sch92} Sch\"{u}ssler, M. \& Solanki, S.K., 1992, \aap, 264, L13
\bibitem[Sch\"{u}ssler et al. (1996)]{Sch96} Sch\"{u}ssler, M., Caligari, P., Ferriz-Mas, A., Solanki, S.K. \& Stix, M., 1996, \aap, 314, 503
\bibitem[Shakhovskaya (1974)]{Sha74} Shakhovskaya, N. I., 1974, IBVS, No.897, 1
\bibitem[Spiesman \& Hawley (1986)]{Spi86} Spiesman, W. J. \& Hawley, S. L., 1986, \aj, 92, 664
\bibitem[Stellingwerf (1978)]{Ste78} Stellingwerf, R.F., 1978, \apj, 224, 935
\bibitem[Torres \& Ferraz Mello (1973)]{Fer73} Torres, C. A. O. \& Ferraz Mello, S., 1973, \aap, 27, 231
\bibitem[Veeder (1974)]{Vee74} Veeder, G. J., 1974, \aj, 79, 702V 
\bibitem[Vogt (1975)]{Vog75} Vogt, S. S., 1975, \apj, 199, 418
\bibitem[Wilson (1954)]{Wil54} Wilson, R. H., Jr., 1954, \aj, 59, 132
\bibitem[Wilson (1978)]{Wil78} Wilson, O.C., 1978, \apj, 226, 379
\end{thebibliography}
\end{document}